\documentclass[final,3p,times]{elsarticle}
\usepackage{graphicx}
\usepackage{wrapfig}
\usepackage{amssymb}
\usepackage{mathrsfs}
\usepackage{amsmath}
\usepackage{comment}
\usepackage{amsfonts}
\usepackage{epsf,color,colordvi,pifont}
\usepackage{lineno}
\usepackage{nicefrac}
\usepackage[notref,notcite]{showkeys} 

\DeclareFontFamily{OT1}{pzc}{}
\DeclareFontShape{OT1}{pzc}{m}{it}{<-> s * [1.10] pzcmi7t}{}
\DeclareMathAlphabet{\mathpzc}{OT1}{pzc}{m}{it}

\usepackage{amsmath}

\journal{Annals of physics}
\begin{document}
\begin{frontmatter}
\title{Searching for  minicharged particles via birefringence, dichroism and Raman spectroscopy of the vacuum polarized by a high-intensity  laser wave}

\author{S.  Villalba-Ch\'avez}
\ead{selym@tp1.uni-duesseldorf.de}
\author{C.  M\"{u}ller}
\ead{c.mueller@tp1.uni-duesseldorf.de}
\address{Institut f\"{u}r Theoretische Physik I, Heinrich Heine Universit\"{a}t D\"{u}sseldorf\\ Universit\"{a}tsstr. 1, 40225 D\"{u}sseldorf, Germany}
\begin{abstract}
Absorption and dispersion of  probe photons  in the field  of a high-intensity circularly polarized  laser wave  are investigated.
The optical theorem is applied for determining the absorption coefficients  in terms of  the imaginary part of the
vacuum polarization tensor. Compact  expressions  for the vacuum refraction indices and  the photon absorption coefficients
are  obtained  in various asymptotic regimes of interest. The outcomes of this  analysis  reveal  that,  far from the  region relatively close to the
threshold of  the two-photon reaction, the  birefringence and dichroism of the vacuum are small and, in some cases, strongly suppressed.  On the contrary,
in a vicinity of the region in which the photo-production of a pair occurs, these optical properties are  manifest with
lasers of moderate intensities. We take advantage of  such a property  in the search of    minicharged particles by considering
high-precision polarimetric experiments. In addition,  Raman-like electromagnetic waves  resulting  from the
inelastic part  of  the vacuum polarization tensor  are  suggested  as an alternative form  for finding  exclusion limits
on  these hypothetical charge carriers.  The envisaged  parameters of upcoming high-intensity laser facilities are used
for establishing upper bounds on the minicharged particles.
\end{abstract}
\end{frontmatter}

\section{Introduction}

Investigating the  frontiers of the Standard Model (SM)  is  a  fundamental  issue  in elementary  particle physics.
Despite the successes of the minimal $SU(3)\times SU(2)\times U(1)$ gauge  group as unified description of the strong and  electro-weak
interaction, there  still remain  a variety of nontrivial  issues whose solutions  often  demand physics beyond the SM.
The absence of a satisfactory explanation for the large number  of free parameters as well as the hierarchy and naturalness problems
to which they are subject constitute clear examples of unresolved questions.  These  seem to be  direct consequences of
dealing with an effective formulation  rather than a fundamental theory where,
among other issues,   gravity is  conceptually reconciliated with  the remaining  interactions. In connection, several SM extensions have
been put forward. At energies much above the typical SM scale--specified by the  mass of the $W_\pm-$bosons--supersymmetric versions and string theory are
likely to occur \cite{Weinberg:2000,polchinski}.  In contrast, below the scale given  by the electron mass $m$, the promising candidates introduce
weakly interacting sectors  \cite{Jaeckel:2010ni,Redondo:2010dp,Gies:2007ua,Gies:2008wv}, sometimes,  with  additional  $U(1)$  invariance often
 resulting from  string compactifications \cite{Witten:1984dg,Lebedev:2009ag}.   As a consequence, hypothetical  paraphotons \cite{Okun:1982xi,Ahlers:2007rd,Ahlers:2007qf,Goodsell:2009xc,Dudas:2012pb}--kinematically  mixed with
the  visible $U(1)$ sector--are promoted and  light particles with tiny  fractions of the electron charge are suspected to occur
in nature \cite{Holdom:1985ag,Masso:2006gc,Gies:2006ca,Jaeckel:2009dh,Bruemmer:2009ky}.  Models  which  rely  on   these kind of minicharged
particles (MCPs) are of paramount importance   in contemporary  physics since they constitute by themself ideal   frameworks  for probing
the validity of the SM,  the theoretical conception of magnetic monopoles and, therefore,  the highly nontrivial principle of charge quantization.

Over  the last decade there have been considerable experimental efforts toward  the exploration of the low energy  frontiers of particle
physics, mainly through the unconventional properties of  the unstable and nonlinear  vacuum of Quantum Electrodynamics  (QED). Indeed,
polarimetric experiments  with unprecedented  levels of sensitivity  represent a powerful tool for finding out  stringent  constraints on
the parameters associated with  weakly interacting  particles \cite{Cameron:1993mr,Zavattini:2007ee,BMVreport,Chen:2006cd}. The reasons for
using   these optical setups  follows from  a  hypothetical coupling   between the  MCPs and  a constant magnetic field.  In such a context,
the interaction  would induce  modifications on  the  dichroic and birefringent properties of the  vacuum \cite{Dittrich,adler,shabad4,VillalbaChavez:2012ea,Chavez:2009ia,Hattori},
a fact which constitutes  a  potential  signal of their existences. Also high-precision photon regenerative experiments  have been carried
out in several collaborations \cite{Chou:2007zzc,Steffen:2009sc,Afanasev:2008jt,Pugnat:2007nu,Robilliard:2007bq,Fouche:2008jk,Ehret:2010mh,Ehret:2009sq}.
Most of them relied on  a ``Light Shining Through a Wall'' setup  \cite{VanBibber:1987rq,Adler:2008gk,Arias:2010bh,Dobrich:2012sw,Dobrich:2012jd}, where
the  photon oscillation into a  weakly interacting particle  allows for traversing  a   photon blocker  barrier  and,  eventually,   its regeneration
behind the wall.  In both  types of experiments a tiny effect due to the MCPs is  expected,  but   they might  be more manifest  by increasing
the magnetic field  strength and its spatial extension. Nowadays, it is not  a big issue  to  extend  the effective interaction region up to a
few kilometers  by using mirrors of extremely high reflectivity. In contrast, the  attainable  magnetic field strengths still remain  nine
orders of  magnitude smaller than  the critical one of QED [$B_c=4.42\times10^{13}$ G], near of which, the upper bounds on MCPs are expected
to be quite stringent.

With the progressive  increasing of the available  intensity, laser technology is  becoming  a competitive source of strong fields, valuable
for the search of MCPs. Projects such as  the Extreme Light Infrastructure (ELI) \cite{ELI} and the Exawatt Center for Extreme Light Studies  (XCELS)
\cite{xcels}  are being  designed to reach fields of  about two orders of magnitude   below  the critical one in ultra-short  pulses of temporal
lengths of the order of   $\tau\approx 10\ \rm fs$. Hence, the prospect of finding  stringent  limits  on the MCPs by using high-intensity
lasers   is certainly enticing. Obviously, an essential step in this direction is achieved by knowing  the expressions of the vacuum polarization
tensor in the field of a plane-wave. Although  these were derived a long time ago \cite{baier,Mitter}, up to now their main essential consequence
considered in a realistic context remains the production of electron-positron pairs by a photon--also known as the Breit-Wheeler reaction--\cite{Breit:1934zz,Reiss1962,narozhnyi,Titov,krajewska,VillalbaChavez:2012bb},
and in the Coulomb field of a  nucleus, i.e., the Bethe-Heitler phenomenon \cite{Milstein:2006zz,Dipiazza:2010zz}. So far, the production rates of
the latter processes have not found a direct application in the search of MCPs. In contrast,  the optical  properties  of a polarized QED vacuum
are those which might  provide  sensitive  insights on the parameters associated with these hypothetical charge carriers. Previous considerations
on the optical nature  of the vacuum, polarized  by a circularly polarized laser,  were partially developed in Ref.~\cite{Mitter}, where a  numerical
assessment of the photon  absorption coefficients and the vacuum refraction indices was undertaken in regions different from the strong field regime.

In contrast to this numerical study, we determine  in the present work  analytical  expressions  for the vacuum refraction indices and  the photon absorption
coefficients  in various asymptotic domains,  including the one related to the strong field domain. Besides, the study of these optical quantities  is extended
to the framework of scalar QED since the spinless realization of MCPs is not discarded  \cite{Ahlers:2006iz,zavattini}.  We show that, far from the threshold of
two-photon reaction--and independently of the nature of the charge carriers--the  birefringence and dichroism properties of the vacuum are small, and  in some regions,
strongly suppressed. On the contrary,  in a vicinity of the first  photo-production threshold, the birefringence and dichroism of the vacuum are considerably more
pronounced. Both phenomena are closely connected with the chiral activity of the ``medium'' and--according to our results--could be observed even at
intensities available today. We take advantage of  such a property  in the search of    MCPs  by considering high-precision polarimetric techniques.  Because of the relative weakness of  the
aforementioned phenomenon in the strong field regime,  we look for   an observable different from  the ellipticity and  the rotation of the polarization plane,
both frequently considered in the case where a dipole magnet  drives the polarization of the vacuum.  In fact,  we  take   into account Raman-like electromagnetic
waves  arising   from  the inelastic interaction in the photon-photon scattering. The associated  spectroscopy techniques are  then suggested  for  probing the quantum
vacuum but also for determining   upper bounds  on the   parameters intrinsically associated  with the MCPs.

\section{Photon scattering  in a circularly polarized wave}

\subsection{The polarization tensor, its tensorial structure and form factors}

We are motivated to investigate the effects induced by   hypothetical  particles characterized by a mass $m_\epsilon$ and  a tiny fraction of the electron charge
$\mathpzc{q}_\epsilon\equiv\epsilon\vert e\vert$.  As long as the energy scale remains within the phenomenological  limits of  QED and  its  fundamental principles
are preserved, the  consequences associated with the existence of this  sort of MCPs  would  not differ from those emerging in a pure QED context.
Conceptually,  one can then  investigate the related phenomenology  from  the already  known  expressions,  with    the electron  parameters $(e,\ m)$   substituted
by the respective quantities associated with an MCP $(\mathpzc{q}_\epsilon,\ m_\epsilon).$  Besides, by probing  the nonlinear and unstable  properties
of  the  $\mathpzc{q}_\epsilon^{+} \mathpzc{q}_\epsilon^{-}$ vacuum in  an  external background field,  stringent limits on  the presumable  smallness of the  unknown
parameters $\epsilon$ and $m_\epsilon$ can be obtained. This constitutes a motivation for  studying the dispersive and absorptive processes intrinsically connected to
the vacuum polarization effects. The latter become  primarily manifest in the generating functional of one-particle irreducible diagrams
\begin{equation}\label{effectiveaction}
\Gamma=\frac{1}{8\pi}\int d^4x\ d^4x^\prime\left\{a_\mu(x)\left[\left(\square \mathpzc{g}^{\mu\nu}-\partial^\mu\partial^\nu\right)\delta^4(x-x^\prime)+\Pi^{\mu\nu}(x,x^\prime)\right]a_\nu(x^\prime)\right\}+\ldots
\end{equation}where the metric tensor reads $\mathpzc{g}_{\mu\nu}=\mathrm{diag}(+1,-1,-1,-1)$,  $\square\equiv\partial_\mu\partial^\mu=\partial^2/\partial t^2-\nabla^2$ and the abbreviation   $+\ldots$ stands for
higher order terms in  the small-amplitude electromagnetic wave  $a_\mu(x)$. The respective  Dyson-Schwinger equation \cite{Dyson:1949ha,Schwinger:1951ex1,Schwinger:1951ex2,fradkin,Alkofer:2000wg},
written up to linear order  in $a_\mu(x)$, has  the form
\begin{eqnarray}
\square a_\mu(x)+\int d^4x^\prime\Pi_{\mu\nu}(x,x^\prime)a^{\nu}(x^\prime)=0  \label{sdpmBF}
\end{eqnarray}   provided  that  $a_\mu(x)$  is    chosen in the Lorenz gauge $\partial a =0.$  While the first term in Eq.~(\ref{sdpmBF}) is the classical Maxwell contribution,
the second one is  responsible for the interaction of photons with the external background  field.  Such an interaction  is mediated by   the virtual  minicharged carriers
and  encompassed  in the vacuum  polarization tensor  $\Pi_{\mu\nu}(x,x^\prime)$. It is through this object that  the gauge sector of QED
acquires a  dependence on the strong field  of the wave. The latter is taken hereafter  as a circularly polarized
monochromatic plane-wave\footnote{From now on ``natural'' and Gaussian  units  $c=\hbar=4\pi\epsilon_0=1$ are used.}
\begin{eqnarray}
\mathscr{A}^{\mu}(x)=\mathpzc{a}_1^{\mu}\cos(\varkappa x)+\mathpzc{a}_2^{\mu}\sin(\varkappa x)\quad  \mathrm{with}\quad  \mathpzc{a}_1\mathpzc{a}_2=0,\quad \varkappa^2=0,\quad \mathpzc{a}_1^2=\mathpzc{a}_2^2\equiv  \mathpzc{a}^2,
\label{externalF}
\end{eqnarray}and   specialized  in   the Lorenz gauge $\partial  \mathscr{A}=0$ as well. This condition implies that  the wave four-vector  $\varkappa^{\mu}=(\varkappa^0,\pmb{\varkappa})$
and  the  constant  polarization vectors  $\mathpzc{a}_i^\mu$  (with $i=1,2$) satisfy the constraints  $\varkappa \mathpzc{a}_i=0.$  With these details in mind,  we Fourier transform Eq.~(\ref{sdpmBF})
\begin{equation}
k_1^2 a_\mu(k_1)-\int \frac{d^4k_2}{(2\pi)^4} \Pi_{\mu\nu}(k_1,k_2)a^\nu(k_2)=0,\label{dsequation}
\end{equation}and consider the one-loop approximation in $\Pi_{\mu\nu}(k_1,k_2)$, in which  case the polarization tensor acquires
the following structure (for details we refer the reader to Ref.~\cite{baier})
\begin{eqnarray}\label{ptcircular}
\Pi^{\mu\nu}(k_1,k_2)=(2\pi)^4\delta^4(k_1-k_2)\Pi^{\mu\nu}_0(k_2)+(2\pi)^4\delta^4(k_1-k_2+2\varkappa)\Pi^{\mu\nu}_{+}(k_2)+(2\pi)^4\delta^4(k_1-k_2-2\varkappa)\Pi^{\mu\nu}_{-}(k_2).
\end{eqnarray}The tensorial objects involved in this expression can be expanded in terms of the following Lorentz covariant vectors
\begin{eqnarray}\label{vectorialbasisbaier}
\Lambda_{\pm}^\mu(k)=-\frac{\left(\mathscr{F}_{1}^{\mu\nu}\pm i\mathscr{F}_{2}^{\mu\nu}\right)k_\nu}{(k\varkappa)\left(-\mathpzc{a}^2\right)^{\nicefrac{1}{2}}},\quad \Lambda_3^\mu(k)=\frac{\varkappa^\mu k^2-k^{\mu} (k\varkappa)}{(k\varkappa)\left(k^2\right)^{\nicefrac{1}{2}}}
\end{eqnarray} where  $\mathscr{F}^{\mu\nu}_{i}=\varkappa^\mu\mathpzc{a}^\nu_{i}-\varkappa^\nu\mathpzc{a}^\mu_i$  denotes  the field strengths tensor
associated with each external field mode $(i=1,2)$.   We emphasize that Eq.~\eqref{vectorialbasisbaier} does not depend on which choice of $k$ is taken since the
difference between $k_1$ and $k_2$ is proportional  to $\varkappa$. Note that  $\Lambda_\pm$  are eigenstates of the helicity operator   subject to  the normalization conditions
$\Lambda_+\Lambda_-=-2$ with  $\Lambda_\pm\Lambda_\pm=0.$ Besides, we also find that $\Lambda_3\Lambda_3=-1$ with $\Lambda_\pm\Lambda_3=0$.
With  Eq. (\ref{vectorialbasisbaier}) in mind,  the part in Eq. (\ref{ptcircular}) responsible for the elastic scattering can be written as
\begin{eqnarray}\label{elasticpart}
\Pi^{\mu\nu}_0(k_1)= \frac{1}{2}(\pi_3+i\pi_1)\Lambda^\mu_+\Lambda^\nu_-+ \frac{1}{2}(\pi_3-i\pi_1)\Lambda^\mu_-\Lambda^\nu_++\pi_5\Lambda_3^\mu\Lambda_3^\nu,
\end{eqnarray}whereas the tensors associated with  inelastic scattering turn out to be $\Pi^{\mu\nu}_\pm(k_1)= \pi_0\Lambda^\mu_\pm\Lambda^\nu_\pm$.
These inelastic scattered waves emerge as a consequence of  the simultaneous  emission or  absorption of photons of the high-intensity laser wave upon the
scattering event. As a matter of fact,  they  turn out to be shifted to lower or higher values in comparison with  the original monochromatic frequency.
The scattering of light in these latter two cases is analogous to  Raman dispersion in molecular  physics  with $\varkappa$ imitating the vibrational
frequency of the molecules. Its relevance   will be analyzed separately in Sec. \ref{rammanwavespoltensor}.

It is worth  emphasizing that,   owing to  Eq. (\ref{vectorialbasisbaier}), the polarization tensor satisfies not only  the fundamental
principles  of charge conjugation, time reversal and parity symmetry but also the  gauge invariance properties of the  electromagnetic
interaction. The form factors $\pi_i$ involved in Eq.~(\ref{elasticpart})  turn out to be  twofold parametric
integrals\footnote{The
form factors defined  in Eq. (\ref{formfactors}) and (\ref{g3}) are in correspondence with those obtained by Ba\u{\i}er, Mil'shte\u{\i}n and Strakhovenko in Ref. \cite{baier} -
according to the renaming rule $\pi_i\Leftrightarrow\alpha_i$. }
\begin{equation}\label{formfactors}
 \pi_i(\lambda_\epsilon,\xi_\epsilon)=-\frac{\alpha_\epsilon}{2\pi} m_\epsilon^2\int_{-1}^{1}dv \int_0^\infty \frac{d\rho}{\rho}e^{-\frac{2i\rho}{\vert\lambda_\epsilon\vert(1-v^2)}\left[1+2A\xi_\epsilon^2-\frac{k_1k_2\left(1-v^2\right)}{4m_\epsilon^2}\right]}\Omega_i.
\end{equation}with
\begin{eqnarray}
\begin{array}{c}\displaystyle
\Omega_1^{(0)}=2\xi_\epsilon^2\rho A_0 \mathrm{sign}\left[\lambda_\epsilon\right],\qquad \Omega_1^{(\frac{1}{2})}=4\xi_\epsilon^2\rho A_0\frac{1+v^2}{1-v^2}\mathrm{sign}\left[\lambda_\epsilon\right]\\ \\ \Omega_5^{(0)}=-\frac{k_1^2}{4m_\epsilon^2}v^2\left(1-e^{iy}\right),\qquad \Omega_5^{(\frac{1}{2})}=-\frac{k_1^2}{2m_\epsilon^2}(1-v^2)\left(1-e^{iy}\right),
\\ \\ \displaystyle \Omega_3^{(0)}=\xi_\epsilon^2\sin^2\left(\rho\right)+\frac{1}{2}\left[1-\frac{k_1^2}{4m_\epsilon^2}(1-v^2)\right]\left(1-e^{iy}\right), \quad\Omega_3^{(\frac{1}{2})}=2\xi_\epsilon^2\sin^2\left(\rho\right)\frac{1+v^2}{1-v^2}-\left[1+\frac{k_1^2}{4m_\epsilon^2}(1+v^2)\right]\left(1-e^{iy}\right).
 \end{array}\label{g3}
\end{eqnarray} While the  upper index $0$ denotes the quantities  coming out  from a loop of scalar particles,  the upper index  $\frac{1}{2}$ refers to the case where the  spin-$\frac{1}{2}$
representation  mediates the interaction. Here  $\alpha_\epsilon\equiv \epsilon^2 e^2=\epsilon^2/137$ denotes the fine structure constant relative to the MCPs
and  $\xi_\epsilon^2=-\epsilon^2 e^2\mathpzc{a}^2/m_\epsilon^2$. Other functions and parameters,   contained   in  these expressions,  are given by
\begin{eqnarray}
\begin{array}{c}\displaystyle
A=\frac{1}{2}\left(1-\frac{\sin^2(\rho)}{\rho^2}\right),\quad  A_0=\frac{1}{2}\left(\frac{\sin^2(\rho)}{\rho^2}-\frac{\sin(2\rho)}{2\rho}\right),\quad A_1=A+2A_0,\quad y=\frac{4\rho\xi_\epsilon^2 A}{\vert\lambda_\epsilon\vert(1-v^2)},\quad \lambda_\epsilon=\frac{\varkappa k}{2m_\epsilon^2}.
\end{array}
\label{parameters}
\end{eqnarray}

Because of  the high-oscillatory behavior  of the functions $\Omega_i$,  an exact analytical evaluation of the  form factors $\pi_i$  cannot be carried out.
Instead,  we shall focus ourselves on their  asymptotic features in various limits of interest with respect to the parameters $\lambda_\epsilon$ and $\xi_\epsilon$.

\subsection{Dispersion and absorption of small-amplitude electromagnetic waves. General considerations.}

The transversal small-amplitude electromagnetic wave, solution  of  Eq.~(\ref{dsequation}), can be expressed  as a
superposition  of two  different helicity modes
\begin{equation}\label{chiralityexpansio}
a^{\mu}(k)=f_+(k)\Lambda^\mu_++f_-(k)\Lambda^\mu_-.
\end{equation}In order to find their propagation laws  we  substitute  Eq. (\ref{chiralityexpansio}) into Eq.~(\ref{dsequation})
by inserting, in addition, Eq.~(\ref{ptcircular}) and Eq.~(\ref{elasticpart}). The resulting equation
is   projected onto the  $\Lambda^\mu_\pm$  afterwards. The described procedure  provides the following  system of equations
\begin{equation}\label{eigenproblem}
\pmb{\mathpzc{G}}^{(i)}(k)\pmb{z}^{(i)}(k)=0\quad \mathrm{with}\quad i=1,2.
\end{equation}Here the involved quantities are defined as follows:
\begin{equation}\label{dispr1}
\pmb{\mathpzc{G}}^{(1)}(k)=\left[
\begin{array}{ccc}
k^2+\pi_3+i\pi_1 &2\pi_0^+\\
2\pi_0 & (k+2\varkappa)^2+\pi_3^+-i\pi_1^+
\end{array}\right],\quad \pmb{\mathpzc{G}}^{(2)}(k)=\left[
\begin{array}{ccc}
(k-2\varkappa)^2+\pi_3^-+i\pi_1^- &2\pi_0\\
 2\pi_0^- & k^2+\pi_3-i\pi_1
\end{array}\right],
\end{equation}
\begin{equation}
\pmb{z}^{(1)} =
\left[\begin{array}{c} f_+(k)\\ f_-(k+2\varkappa)
\end{array}\right],\quad \pmb{z}^{(2)} =
\left[\begin{array}{c}
 f_+(k-2\varkappa)\\ f_-(k)
\end{array}\right].\label{statesflavornomass}
\end{equation}Note that the form factors  having  an upper  index $\pm$ must be evaluated at $k\to k\pm2\varkappa.$
Because of this fact, both eigenproblems turn out to be  correlated, i.e., $\pmb{\mathpzc{G}}^{(2)}(k)=\pmb{\mathpzc{G}}^{(1)}(k-2\varkappa)$.
Of course, the dispersion  relations emerge whenever  the determinant of $\pmb{\mathpzc{G}}^{(i)}(k)$  vanishes identically. Its solutions can
be determined by analytical procedures. However,  we will consider  the situation  in which  the polarization effects do  not modify dramatically
the usual  photon dispersion law $\omega=\vert\pmb{k}\vert$. Thereby  only leading order corrections in $\alpha_\epsilon$ will be taken into
account.  Guided by this approximation the relevant dispersion equations for $f_\pm(k)$ turn out to be
\begin{equation}
k^2+\pi_3\pm i\pi_1\simeq0,\label{dispersion}
\end{equation}where  the contribution resulting from the off-diagonal terms  in Eq.~(\ref{dispr1}) has been neglected since it  provides a correction
smaller by a factor  $\alpha_\epsilon$.

The  polarization tensor $\Pi_{\mu\nu}$ is, in  general, a non-hermitian object. In correspondence, its  form factors contain  real and
imaginary contributions $\pi_i=\mathrm{Re}\ \pi_i+i\ \mathrm{Im}\ \pi_i$. The respective dispersion relations, solutions of Eq.~(\ref{dispersion}),  must
be  complex functions as well, i.e., $\omega_\pm=\mathrm{Re}\ \omega_\pm+i\mathrm{Im}\ \omega_\pm$. While the real part describes the dispersive phenomenon, the imaginary
contribution  provides the absorption coefficient $\kappa_\pm\equiv-\mathrm{Im}\ \omega_\pm$ of  mode-$\pm$ photon.  This  analysis, together with
the definition of  the vacuum refractive index $n_\pm=\vert\pmb{k}\vert/\mathrm{Re}\ \omega_\pm$, allows us to  establish the relations
\begin{equation}
n_\pm^2-1=\left.\frac{\mathrm{Re}\ \pi_3\mp\mathrm{Im}\ \pi_1}{\mathrm{Re}\ \omega_\pm^2}\right\vert_{k^2=0}\qquad
\mathrm{and}\qquad \kappa_\pm=\left.\frac{\mathrm{Im}\ \pi_3\pm\mathrm{Re}\ \pi_1}{2\mathrm{Re}\ \omega_\pm}\right\vert_{k^2=0}.\label{refractioninde}
\end{equation}  Observe that
the vacuum occupied by the field of the wave [Eq.~(\ref{externalF})] is  birefringent whenever $\mathrm{Im}\ \pi_1$ does not vanish identically. We should also mention
at this point that the sum of the  absorption coefficients coincides with the rate of the photo-production of a $\mathpzc{q}_\epsilon^+\mathpzc{q}_\epsilon^-$ pair
in a  circularly polarized wave averaged over the photon polarization states \cite{VillalbaChavez:2012bb}. The latter statement is expected since the imaginary part
of the polarization tensor is associated with the probability of the pair creation through the optical theorem. Indeed, within the accuracy to the second order with
respect to the radiative corrections, the total creation rate  of a $\mathpzc{q}_\epsilon^+ \mathpzc{q}_\epsilon^-$ pair  from a photon  with polarization
$\mathpzc{e}_\ell$ ($\ell=1,2$) turns out to be
\begin{equation}\label{rate}
\Re_\ell=\frac{\mathpzc{e}^{\mu *}_\ell\mathpzc{e}_\ell^\nu}{\omega}\mathrm{Im}\ \Pi_{0\mu\nu}(k_1).
\end{equation}As long as the  photon polarizations are  chosen as  $\mathpzc{e}_\pm^\mu=\Lambda_\pm^\mu/2^{\nicefrac{1}{2}}$, the expression above reduces to
$\Re_\pm=2\kappa_\pm.$  The corresponding  average, on the other hand, turns out to be $\Re=(\Re_++\Re_-)/2=\mathrm{Im}\ \pi_3/\omega$.\footnote{Alternatively, the photo-production rate  can be calculated from the corresponding transition-amplitude in which the exact  nonstationary
solution of the Dirac equation for an electron in the field of the wave  is considered \cite{Reiss1962,narozhnyi}.}  Detailed
calculations of  $\mathrm{Im}\ \pi_3$ may be found in  separate papers (see Refs. \cite{VillalbaChavez:2012bb} and \cite{baierbook}) for the various
limits to be considered in this work. Because of this fact, in the following  we shall be concerned with the determination of the corresponding asymptotic
expressions for the remaining quantities  contained in Eq.~(\ref{refractioninde}).

\section{Elastic absorptive  properties of the quantum  vacuum}

\begin{wrapfigure}{l}{0.47\textwidth}
\includegraphics[width=.47\textwidth]{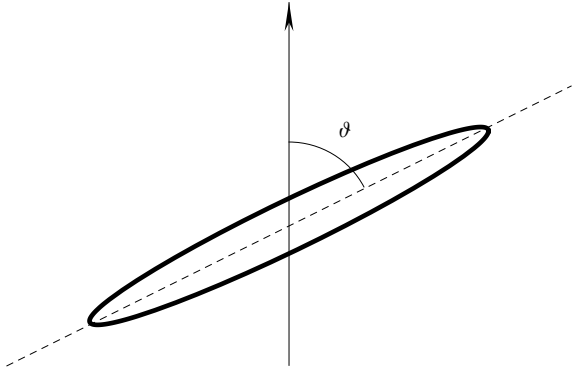}
\caption{\label{fig.000}A schematic representation of the optical effects induced on a probe beam after the interaction with a high-intensity
circularly polarized laser. The vertical axis must be understood as the direction in which the  incoming  monochromatic wave is linearly polarized.
As a consequence of the interaction with the strong field, the outgoing wave turns out to be  elliptically polarized [Eq.~(\ref{rotangle})] with the principal
axis of the ellipse rotated by  a small angle $\vartheta$ [Eq.~(\ref{ellipticity})].}
\end{wrapfigure}

The absorption coefficients  $\kappa_{\pm}$ in Eq.~(\ref{refractioninde})  determine  the decrement of the probe wave-amplitude  due to the
production  of a pair of MCPs.  In order to find  an observable   effect,  we  take  the incoming probe beam  to be a linearly polarized plane
wave. Upon entering the region  occupied by  the external field the probe beam is  decomposed  into its right and left circular-polarized  waves
[see Eq.~(\ref{chiralityexpansio})], which  initially possess equal amplitudes. As a consequence of the vacuum dichroism, the outgoing probe beam  is elliptically
polarized [see  Fig.~(\ref{fig.000})] and the following relation for the  ellipticity $\psi$ is found
\begin{equation}
\sin(2\psi)=\left\vert\frac{e^{-2\kappa_+\tau}-e^{-2\kappa_-\tau}}{e^{-2\kappa_+\tau}+e^{-2\kappa_-\tau}}\right\vert.
\end{equation}Here $\tau$ indicates  the interacting time. Note that, contrary  to  the situation where a constant magnetic field drives the dichroism
\cite{Ahlers:2007rd,Gies:2006ca,Ahlers:2006iz},  $\psi(\tau)$ is  determined here by  the damping factors  $e^{-\kappa_\pm\tau}$ associated with the two
propagating modes. The effect is expected  to be tiny $\psi\ll 1$ and, consequently, the previous expression can be approached to
\begin{equation}
\psi(\tau)\simeq \frac{1}{2}\vert\kappa_--\kappa_+\vert\ \tau.\label{rotangle}
\end{equation}We remark that the last formula is a good approximation only when $\vert\kappa_--\kappa_+\vert\  \tau \ll1$. Incidentally,
Eq.~(\ref{rotangle}) also  applies when an  optically active crystal is studied [for details see \cite{sophocles} and references therein].
This fact  allows then   to establish an analogy between our problem  and  the optics associated with  a  chiral  medium.

In this context, a purely kinematical analysis  proves to be very convenient for forthcoming considerations. To this end we
inspect the energy-momentum conservation associated with an  absorptive process where $n$ photons of the strong wave  are  absorbed in addition to a probe photon.
In the center-of-mass frame, this is given by $k+n\kappa=q_+ +q_-$  where the four-momentum
\[
q_\pm^\mu\equiv(\varepsilon,\pm\pmb{q})=p_\pm^\mu+\frac{m_\epsilon^2\xi_\epsilon^2}{2(\varkappa p_\pm)}\varkappa_\mu\quad \mathrm{with}\quad \quad p_\pm^2=m_\epsilon^2
\]  is  the appropriate  translation generator of the vacuum symmetry group in  an external field \cite{Richard}.
Consequently,  we find the relation  $n k\varkappa=2\varepsilon^2$ with $\varepsilon$ being the laser-dressed energy. Here  the
relative speed between the  final  particle states  turns out to be
\begin{equation}\label{speed}
\vert\pmb{\mathpzc{v}}_{\mathrm{rel}}\vert=\vert\pmb{\mathpzc{v}}_{-}-\pmb{\mathpzc{v}}_{+}\vert=2\mathpzc{v}\quad \mathrm{with}\quad
\mathpzc{v}=\left(1-\frac{n_*}{n}\right)^{\nicefrac{1}{2}}.
\end{equation} Eq. (\ref{speed}) reveals that the photo-production of a $\mathpzc{q}_\epsilon^+\mathpzc{q}_\epsilon^-$ pair  may take place  whenever
the  number of  absorbed photons  of the high-intensity laser wave exceeds the threshold value $n_*=2\mathpzc{m}_{*}^2/k\varkappa.$
In this context,   $\mathpzc{m}_{*}\equiv m_\epsilon(1+\xi_\epsilon^2) ^{\nicefrac{1}{2}}$ must be understood as the effective mass which the
MCPs  acquire  due  to  the   field of the wave.

\subsection{Two-photon reaction and circular dichroism   at  $\xi_\epsilon<1$ \label{twophotonreaction}}

Let us start by determining  $\kappa_{\pm}$ [Eq.~(\ref{refractioninde})] as the intense laser  parameter $\xi_\epsilon < 1$ and   $ n_*\leqslant1$.  Combining
the previous conditions   we obtain  that  the results to be  derived in this subsection do apply  whenever the  inequality $\lambda_\epsilon>2\xi_\epsilon^2$
is fulfilled.  In this case,  the  oscillatory term present in  the exponent of Eq.~(\ref{formfactors}) is   smaller  than  the  remaining contributions.
Consequently,  we can  ignore  it and deal with the following expression
\begin{eqnarray}
&&\pi_1^{(\frac{1}{2})}\simeq-\frac{4\alpha_\epsilon m_\epsilon^2\xi_\epsilon^2}{\pi}\ \int_0^1dv\int_{0}^\infty d\rho e^{\frac{-2i\rho \left(1+\xi_\epsilon^2\right)}{\lambda_\epsilon\left(1-v^2\right)}}
\frac{1+v^2}{1-v^2}A_0\label{scalarprobaborn0}
\end{eqnarray}where the symmetry of the integrand in the variable $v$ was exploited. Note  that for a  photon fulfilling the light cone equation, i.e.,
$k^2=0$,  the parameter $\lambda_\epsilon$ in Eqs.~\eqref{formfactors}
and \eqref{g3} is always nonnegative, $\lambda_\epsilon\geqslant0$. Because of this fact, we have explicitly  taken   in  Eq.~\eqref{scalarprobaborn0}
[and also in the following]  both $\vert\lambda_\epsilon\vert=\lambda_\epsilon$ and  $\mathrm{sign}[\lambda_\epsilon]=1$.

In order to derive an explicit expression of $\mathrm{Re}\ \pi_1^{(\frac{1}{2})}$, we first  integrate by parts the terms containing a
factor proportional to $1/\rho^2$.  The residue theorem  is   applied afterwards. The latter step  requires an  integration  contour   slightly
below the real $\rho$ axis (for  details  we refer the reader to chapter $3$ in  \cite{baierbook}). As a consequence,  we find
\begin{eqnarray}
\mathrm{Re}\ \pi_1^{(\frac{1}{2})}\simeq-\frac{\alpha_\epsilon m_\epsilon^2\xi_\epsilon^2}{2}\ \int_0^1dv\  \Theta\left[1-\frac{n_*}{\left(1-v^2\right)}\right]\frac{1+v^2}{1-v^2}\left\{\frac{1}{2}-\frac{n_*}{\left(1-v^2\right)}\right\}\label{intermetiatedsteofun}
\end{eqnarray}where   $\Theta[x]$ is the unit step function.  The latter  provides  a  cut-off from above in the integral contained in Eq.~(\ref{intermetiatedsteofun}).
In correspondence,  the divergence at  $v=1$ is removed and  the variable $v$ can be integrated out without any  complications. With these details in mind,  we end up with
\begin{eqnarray}\label{repi1fer}
\mathrm{Re}\ \pi_1^{(\frac{1}{2})}\simeq-\frac{ \alpha_\epsilon m_\epsilon^2 \xi_\epsilon^2}{2}\left\{\ln\left(\frac{1+\mathpzc{v}}{1-\mathpzc{v}}\right)-3\mathpzc{v}\right\}\Theta[\mathpzc{v}^2]\label{twophreactgeneral}
\end{eqnarray}where  $\mathpzc{v}=(1-n_*)^{\nicefrac{1}{2}}$ refers to the relative speed where only one  photon of the high-intensity laser wave has been  absorbed. Accordingly,
the photo-production  of a pair of MCPs could take place through a  two-photon reaction $k+\varkappa\to\mathpzc{q}_\epsilon^++\mathpzc{q}_\epsilon^-$.  We combine  this result  with the
$\mathrm{Im}\ \pi^{(\frac{1}{2})}$  previously computed in Appendix E of Ref.~\cite{baierbook} to express the absorption coefficients [Eq.~(\ref{refractioninde})]
of the spinor  QED in the  following form:
\begin{eqnarray}\label{fermionpositivekappa}
&&\kappa_+^{(\frac{1}{2})}=\frac{\alpha_\epsilon m^2_\epsilon \xi_\epsilon^2}{4\omega}\left\{\frac{1-\mathpzc{v}^4}{2(1+\xi_\epsilon^2)}\ln\left(\frac{1+\mathpzc{v}}{1-\mathpzc{v}}\right)+2\mathpzc{v}\left(1-\frac{1-\mathpzc{v}^2}{2(1+\xi_\epsilon^2)}\right)\right\}\Theta[\mathpzc{v}^2], \\ &&\kappa_-^{(\frac{1}{2})}=\frac{\alpha_\epsilon m^2_\epsilon \xi_\epsilon^2}{4\omega}\left\{\left(2+\frac{1-\mathpzc{v}^4}{2(1+\xi_\epsilon^2)}\right)\ln\left(\frac{1+\mathpzc{v}}{1-\mathpzc{v}}\right)-4\mathpzc{v}\left(1+\frac{1-\mathpzc{v}^2}{4(1+\xi_\epsilon^2)}\right)\right\}\Theta[\mathpzc{v}^2].
\end{eqnarray}

Now, the procedure for determining  $\mathrm{Re}\ \pi_1^{(0)}$ shares  certain  similarities  with the previous  case. Indeed, it can be read off  from  Eq.~(\ref{repi1fer}) by
multiplying the latter by $-1/2$, inserting  the coefficient $(1-\mathpzc{v}^2)$ in front of the logarithmic function and removing the factor $3$ in its last term. On the other hand,
the imaginary  part of $\pi_3^{(0)}$ has been recently computed in Ref.~\cite{VillalbaChavez:2012bb}.  With these details in mind, the  resulting absorption coefficients [Eq.~(\ref{refractioninde})]
associated with  scalar   QED  turn out to be
\begin{eqnarray}\label{scalarpositivekappa}
&&\kappa_+^{(0)}=\frac{\alpha_\epsilon m^2_\epsilon \xi_\epsilon^2}{8\omega}\left\{\mathpzc{v}\frac{1-\mathpzc{v}^2}{1+\xi_\epsilon^2}+\left[1-\mathpzc{v}^2-\frac{1-\mathpzc{v}^4}{2(1+\xi_\epsilon^2)}\right]\ln\left(\frac{1+\mathpzc{v}}{1-\mathpzc{v}}\right)\right\}\Theta[\mathpzc{v}^2],\\  &&\kappa_-^{(0)}=\frac{\alpha_\epsilon m^2_\epsilon \xi_\epsilon^2}{8\omega}\left\{\mathpzc{v}\left(2+\frac{1-\mathpzc{v}^2}{1+\xi_\epsilon^2}\right)-\left[1-\mathpzc{v}^2+\frac{1-\mathpzc{v}^4}{2(1+\xi_\epsilon^2)}\right]\ln\left(\frac{1+\mathpzc{v}}{1-\mathpzc{v}}\right)\right\}\Theta[\mathpzc{v}^2].\label{scalarnegativekappa}
\end{eqnarray}Note that the difference between the absorption coefficients coincides with  $\Delta\kappa\equiv\kappa_+-\kappa_-=\mathrm{Re}\ \pi_1/\omega$. This applies whatever
be the nature of the virtual particles involved in the  loop of the  vacuum polarization tensor. The explicit expression for spin$-\frac{1}{2}$ particles is easily read from
Eq.~(\ref{repi1fer}). On the contrary, when  scalar propagators  determine the loop, the difference turns out to be
\begin{equation}\label{deltakappascalar}
\Delta\kappa^{(0)}=\frac{\alpha_\epsilon m_\epsilon^2\xi_\epsilon^2}{4\omega}\left\{(1-\mathpzc{v}^2)\ln\left(\frac{1+\mathpzc{v}}{1-\mathpzc{v}}\right)-\mathpzc{v}\right\}\Theta[\mathpzc{v}^2].
\end{equation}

Let us consider the situation in which the created particles are  ultrarelativistic  [$\mathpzc{v}\sim1$]. In such a limit,  the  absorption coefficients above behave like
\begin{eqnarray}
\kappa_+^{(\frac{1}{2})}\approx \frac{\alpha_\epsilon m_\epsilon^2 \xi_\epsilon^2}{2\omega},\quad \kappa_-^{(\frac{1}{2})}\approx-\frac{\alpha_\epsilon m_\epsilon^2 \xi_\epsilon^2}{2\omega}\left\{\ln\left(\frac{1-\mathpzc{v}}{2}\right)+2\right\},\quad \kappa_+^{(0)}\approx\mathpzc{o}(1-\mathpzc{v}),\ \qquad
\kappa_-^{(0)}\approx\frac{\alpha_\epsilon m_\epsilon^2 \xi_\epsilon^2}{4\omega}.
\end{eqnarray} Likewise,  the respective differences between
the absorption coefficients are  given by
\begin{eqnarray}\label{ultrarelativisticdeltak}
\Delta\kappa^{(\frac{1}{2})}\approx\frac{\alpha_\epsilon m_\epsilon^2 \xi_\epsilon^2}{2\omega}\left\{\ln\left(\frac{1-\mathpzc{v}}{2}\right)+3\right\}\qquad\mathrm{and}\qquad
\Delta\kappa^{(0)}\approx-\frac{\alpha_\epsilon m_\epsilon^2 \xi_\epsilon^2}{4\omega} .
\end{eqnarray} Next, if the particles are created in the center-of-mass frame almost at rest [$\mathpzc{v}\sim0$]  we find that
\begin{eqnarray}
\kappa_+^{(\frac{1}{2})}\approx \frac{\alpha_\epsilon m_\epsilon^2 \xi_\epsilon^2}{2\omega}\mathpzc{v},\qquad
\kappa_-^{(\frac{1}{2})}\approx\mathpzc{o}(\mathpzc{v}^2),\qquad \kappa_+^{(0)}\approx\frac{\alpha_\epsilon m_\epsilon^2 \xi_\epsilon^2}{4\omega}\mathpzc{v},\qquad
\kappa_-^{(0)}\approx\mathpzc{o}(\mathpzc{v}^2).\label{norelativistic}
\end{eqnarray}These results allow  us to approach   $\Delta\kappa\approx\kappa_+$ independently of  the nature of the created  particles. Note, in addition, that
the limiting case where $\xi_\epsilon\ll 1$ corresponds to the Born approximation. The respective   expressions associated with this  limit can be read off from
Eqs.~(\ref{twophreactgeneral})-(\ref{norelativistic}) by  setting  $\xi_\epsilon=0$ in the effective mass $\mathpzc{m}_*$.

\begin{figure}
\includegraphics[width=3in]{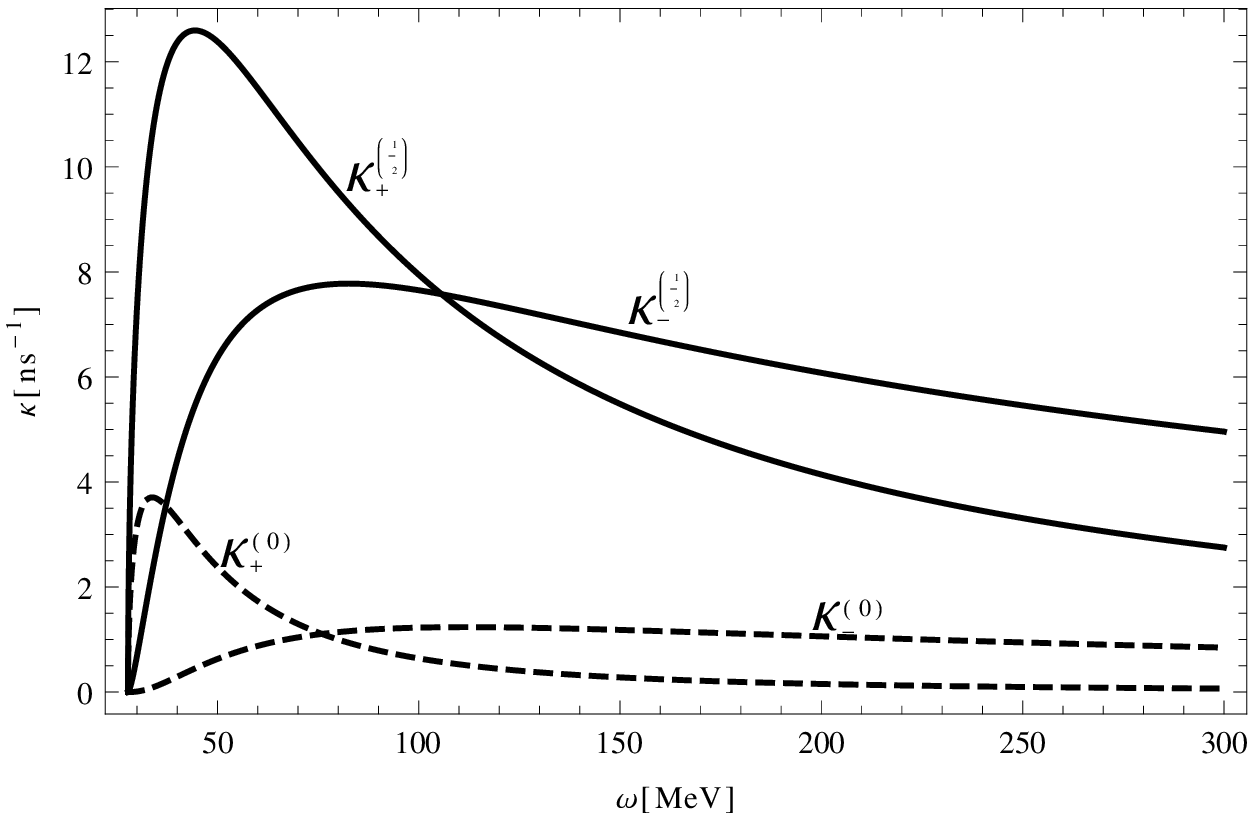}
\includegraphics[width=3.2in]{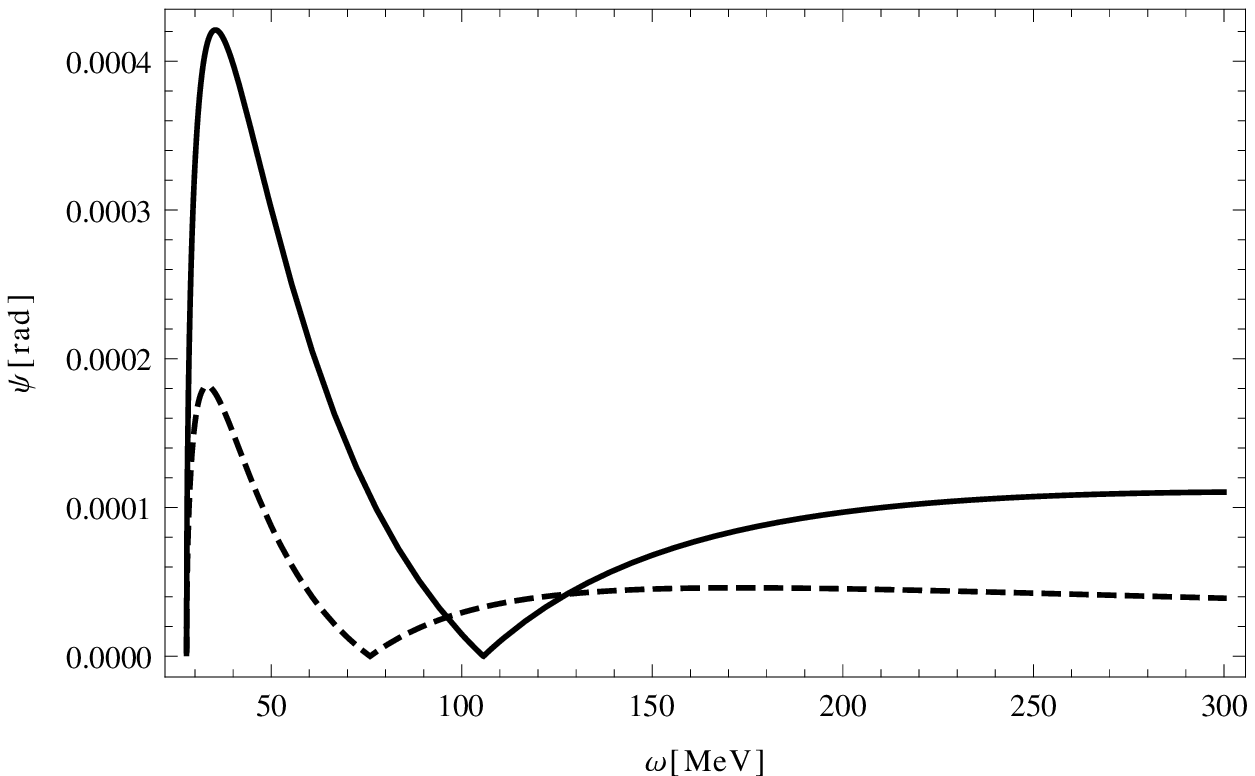}
\caption{\label{fig.001}Dependence of the photon absorption coefficients (left) and the ellipticity  induced by the vacuum dichroism (right) on the frequency of the probe beam.
While the results for spin-$\frac{1}{2}$ particles are represented by  solid curves, those corresponding to the scalar  situation are given by dashed lines. These results  were obtained
by considering  the parameters given in the text: $\xi=7.5\times 10^{-4}$, $\tau\sim 100\;\text{fs}$,  $\varkappa_0=9\ \rm keV$. The particle charge and
mass equal $e$ and $m$, respectively.}
\end{figure}

Formula (\ref{rotangle}), with Eqs.~(\ref{fermionpositivekappa})-(\ref{norelativistic}) included,  is also of relevance  in  a pure QED context.
This is because it   allows us  to determine  the ellipticity induced by the photo-production of an electron-positron pair.   Further analysis  in
this framework requests to set  $\epsilon=1$ and replace   $m_\epsilon$ by the electron mass $m$.  It is  convenient to clarify  that the results
obtained in this way  are valid for $\xi<1$ and  $\lambda>1+\xi^2$ with $\lambda=k\varkappa/(2m^2)$ and $\xi^2=-e^2\mathpzc{a}^2/m^2$. Note besides that, for $\xi\sim 1$  and small values
of $\lambda$  (i.e. $\lambda\sim2$), next-to-leading order terms with respect to $\xi^2/\lambda$ could become relevant.  We assume a strong
laser field with photon energy $\varkappa_0=9\ \rm keV$, intensity parameter $\xi=7.5\times 10^{-4}$ and  temporal length $\tau\sim 100\;\text{fs}$. For this
choice, which is inspired by the x-ray free-electron laser facilities (XFEL) currently under construction at DESY (Hamburg, Germany) and SLAC (Standford, USA),  the
first  Born approximation can be used.   With this set  of parameters in mind, the photo-production of an electron-positron pair  takes place if the probe beam
has a frequency  $\omega\geqslant 27.8\ \rm MeV$, assuming  a head-on collision geometry.  When Dirac particles  are created, the   induced ellipticity is
maximized  at  $\omega\approx 35.3\ \rm MeV$. For further information, we refer  the reader to Fig.~(\ref{fig.001}), where  the dependence
of $\psi$  with respect to $\omega$  is displayed.

\subsection{Extinction of the vacuum dichroism in the limit  $\xi_\epsilon\gg1$\label{quasi}}

We wish to   find  out the leading order terms  of the photon absorption coefficients   $\kappa_\pm$ [Eq.~(\ref{refractioninde})] as $\xi_\epsilon\gg1$.
In this case, the  photo-production rate  of a $\mathpzc{q}^+_\epsilon\mathpzc{q}^-_\epsilon$ pair  is  independent of the frequency
of the external laser  beam $\varkappa_0$  and  coincides with the rate arising in the constant crossed field configuration
\cite{VillalbaChavez:2012bb}.   To find the corresponding difference between the absorption coefficients it is convenient to  carry
out the change of  variable $u=(1-v^2)^{-1}$ in Eq.~(\ref{formfactors}). This leads to express
\begin{eqnarray}\label{modified}
\Delta\kappa^{(\frac{1}{2})}(\lambda_\epsilon,\xi_\epsilon)=\kappa_+^{(\frac{1}{2})}-\kappa_-^{(\frac{1}{2})}=\frac{\mathrm{Re}\ \pi_1}{\omega}=-\frac{2\alpha_\epsilon m_\epsilon^2 \xi_\epsilon^2}{\pi \omega}\int_1^{\infty} \frac{du(2u-1)}{2u\sqrt{u(u-1)}}
\int_{-\infty}^\infty d\rho \ A_0 \cos[2 u\eta]
\end{eqnarray}where the abbreviation $\eta=\frac{\rho}{\lambda_\epsilon}\left[1+2\xi_\epsilon^2A\right]$  has been introduced.
Note that, although the resulting integrand  is a  singular function at $u=1$, the integral over this variable  does  not diverge around this value. Besides, since  the integrand  decreases
as  $\sim1/u^2$ at $u\to\infty$  its  main contribution to Eq.~(\ref{modified}) comes from the region  $u\sim1.$  Regarding the  integration over the remaining variable, here  the integrand
falls  off as $\rho\to\pm \infty$ and is, in addition,  a regular  function in $\rho$.   The experience gained in the calculation of  similar integrals (see, for instance, Refs.
\cite{VillalbaChavez:2012bb} and \cite{Milstein:2006zz})  indicates  that the  asymptotic behavior of   Eq.~(\ref{modified}) as $\xi_\epsilon\gg1$  can be determined  by  splitting  the integration
domain  into three regions:
\begin{equation}
\int_{-\infty}^\infty\ d\rho\ldots=\int_{-\infty}^{-\rho_0}\ d\rho\ldots+\int_{-\rho_0}^{\rho_0}\ d\rho\ldots+\int_{\rho_0}^\infty\ d\rho\ldots
 \label{splitting}
\end{equation}where $\rho_0$ denotes a   positive  dimensionless parameter which  fulfills the conditions
\begin{eqnarray}
\xi_\epsilon^{-1}\ll\rho_0\ll1\qquad\mathrm{and}\qquad (\lambda_\epsilon/\xi_\epsilon^2)^{\nicefrac{1}{3}}\ll\rho_0.\label{constraint}
\end{eqnarray}The second condition implies  that the process under consideration requires a high number of photons which are absorbed from the strong field of the wave,
i.e., $n\geqslant n_*\approx\xi_\epsilon^2/\lambda_\epsilon\gg1$ [multi-photon  reaction]. For further  convenience  we denote the integration regions  on the right-hand side of
Eq.~(\ref{splitting}) from left to right as   \emph{lower  region},  \emph{inner  region} and  \emph{upper  region}, respectively.  Observe that
$\vert\rho\vert\leqslant\rho_0\ll1$ within the  inner  integration region. In correspondence, we Taylor expand  $\eta$  and, separately, the remaining part of the integrand.
Next, the  change of  variable $s=\rho\xi_\epsilon$  is performed in the  resulting integral and also in those defined over the lower and upper regions. Afterwards,
the  integration limit $\rho_0\xi_\epsilon$ is extended to infinity. This last step  provides  no contribution from the integral associated with the lower and upper
regions and yields  to approach the total integral over $\rho$  by
\begin{eqnarray}\label{approax}
\int_{-\infty}^{\infty}\ d\rho\ldots\approx\frac{1}{6\xi_\epsilon^3}\int_{-\infty}^{\infty}ds s^2\cos\left[\frac{2u}{\lambda_\epsilon\xi_\epsilon}\left(s+\frac{s^3}{3}\right)\right].
\end{eqnarray}Some comments are in order. First of all,  the last approximation turns out to be   accurate up to terms that decrease  exponentially, like
$\sim(\rho_0\xi_\epsilon)^{-1}\exp\left[-\frac{2u}{3\lambda_\epsilon\xi_\epsilon}(\rho_0\xi_\epsilon)^3\right]$ and faster. Moreover,  once  the
following representation of the Macdonald function is considered:\footnote{Derivation of Eq. (\ref{jeje1}) requires  to combine Eqs.~(3.695.1-2)  in Ref. \cite{Gradshteyn} and differentiate
the resulting expression twice with respect to  $b$.}
\begin{eqnarray}\label{jeje1}
\int_{-\infty}^\infty dy\ y^2 \cos(by +a y^3)=-\frac{2}{9}\frac{b^{\nicefrac{3}{2}}}{a^{\nicefrac{3}{2}}}K_{\nicefrac{1}{3}} \left(\frac{2}{3\sqrt{3}}\frac{b^{\nicefrac{3}{2}}}{a^{\nicefrac{1}{2}}}\right),
\end{eqnarray} Eq.~(\ref{modified}) acquires the following structure:
\begin{equation}\label{scalastrong}
\Delta\kappa^{(\frac{1}{2})}\simeq\frac{\alpha m_\epsilon^2}{3\sqrt{3} \pi\omega\xi_\epsilon}\int_1^{\infty}\frac{du}{u\sqrt{u(u-1)}} K_{\nicefrac{1}{3}}\left(\frac{4u}{3\zeta_\epsilon}\right)(2u-1).
\end{equation}The situation in which spin-$0$ particles are created does not differ too much from the previous  case.  Hence,  the  leading asymptotic  behavior of $\Delta\kappa^{(0)}$ as $\xi_\epsilon\gg1$ turns out to be
\begin{equation}\label{fermiostrong}
\Delta\kappa^{\left(0\right)}\simeq\frac{\alpha m_\epsilon^2}{6\sqrt{3}\pi \omega \xi_\epsilon}\int_1^{\infty}\frac{du}{u\sqrt{u(u-1)}} K_{\nicefrac{1}{3}}\left(\frac{4u}{3\zeta_\epsilon}\right).
\end{equation}
The set of Eqs.~(\ref{scalastrong}) and (\ref{fermiostrong}) involves  the  abbreviation  $\zeta_\epsilon\equiv \lambda_\epsilon\xi_\epsilon=\frac{\epsilon\omega m^2}{2m_\epsilon^3}\frac{E}{E_c}(1-\hat{\pmb{n}}\cdot\hat{\pmb{\varkappa}})$
with  $E_c=m^2/e=1.3 \times 10^{16}\ \rm V/cm$ the critical electric field of QED.  Here  $\hat{\pmb{n}}=\pmb{k}/\vert\pmb{k}\vert$ and $\hat{\pmb{\varkappa}}=\pmb{\varkappa}/\vert\pmb{\varkappa}\vert$
denote the propagation directions  of the  probe and the strong laser field, respectively. Note that    Eqs.~(\ref{scalastrong}) and (\ref{fermiostrong}) are structurally  similar to the respective expressions
of the photo-production rates  $\Re$  \cite{baierbook,VillalbaChavez:2012bb}.   It is noticeable, however,  that in contrast to the latter,  they are suppressed  by  a  factor  $\sim1/\xi_\epsilon$
which, in addition,  depends on  the frequency of the high-intensity laser. Hence, $\Delta \kappa$ can be understood as  a term which is sensitive to the properties of the strong wave. This kind of
dependence also emerges  in  $\Re$ when  corrections,  next-to-leading  order,  are  taken into account \cite{baierbook}. In order to evaluate the role  of $\Delta\kappa$
within the absorption coefficients, it is convenient to express them  as  $\kappa_\pm=(\Re\pm\Delta\kappa)/2$.  This makes evident that   $\Delta\kappa$  acts  as a small  correction, too. As a
consequence,  the production of pairs  is equally plausible  in either of  the two propagating modes, leading to approach $\kappa_\pm\approx\Re/2\pm\mathpzc{o}\left(\xi_\epsilon^{-1}\right)$.
Here we do not present any picture of the rates $\Re$ because a numerical assessment of  this issue has been recently carried out in Ref. \cite{spin}. Instead, we just emphasize that,  with the
increasing of the intensity of the strong background  laser, the vacuum becomes less and less dichroic to a linearly polarized probe beam, contrary to what occurs in a vacuum polarized by a constant crossed field.

This different behavior is closely connected to the  invariance properties  of each problem. In a  constant crossed field configuration, the vacuum behaves like a biaxial medium and its  symmetry  is no longer
described by  Poincar\'e's group. Instead,  a subgroup of it maps  the actual invariance of the Minkowski space occupied by the external field.  It is  the vacuum polarization  tensor $\Pi_{\mu\nu}$
 which  incorporates  this anisotropy into the gauge sector of QED [Eq.~(\ref{effectiveaction})]. Therefore, the  problem associated with the photon propagation is
no longer  degenerated in  the energy  since the physical  degrees of freedom are  described by birefringent states [for details we refer the reader to Ref. \cite{VillalbaChavez:2012ea}]. Consequently,
the helicity is no longer necessary for labeling the one-particle state. The situation  is quite different in the field of  a circularly polarized  monochromatic plane wave. Here the quantum vacuum
behaves like an anisotropic chiral medium and is   invariant with respect to the following operation: translation by an arbitrary vector $\beta^\mu$ followed by a spatial rotation about the direction of
the wave propagation $\hat{\pmb{\varkappa}}$ of the field by an  angle $\varkappa\beta$ \cite{Mitter,Richard}. Clearly, in the limit $\xi_\epsilon\gg1$ the independence of the high-intensity
laser frequency renders the problem quasi-static with respect to the external field. This means that, in the interaction, the probe beam does not perceive the rotation of  the strong  field of the wave and the
external field seems to be--in average--isotropically distributed in the vacuum. This ``new''  isotropy of  the  spacetime causes the  photon propagation  problem to be quasi-degenerated in the energy.
Hence, the physical modes that emerge from  the interaction can be described  approximately by  monochromatic waves  with opposite helicity.

It is  convenient to remark that the suppression of $\Delta\kappa$ cannot be  compensated  by the dependence on $\xi_\epsilon$  present in the  Macdonald functions
of Eqs.~(\ref{scalastrong}) and (\ref{fermiostrong}). This becomes   manifest  as soon  as the main asymptotic  behaviors of  $\Delta\kappa$ are  taken into account. In order to show the latter
we consider first  the situation  in which  $\zeta_\epsilon\gg1$. Consistency   with our original  condition ($\xi_\epsilon\gg1$) requires to restrict the parameter $\lambda_\epsilon$  to
values with  $\lambda_\epsilon\gg1/\xi_\epsilon$.  Applying  the small-argument  behavior of the functions  $K_\nu(z)\sim\frac{\Gamma(\nu)}{2}\left(\frac{2}{z}\right)^\nu$  \cite{Gradshteyn}
 we  find
\begin{eqnarray}\label{largekappazeta}
\Delta\kappa^{(\frac{1}{2})}\approx\frac{\alpha_\epsilon m_\epsilon^2\lambda_\epsilon}{3\sqrt{3\pi}\omega\zeta_\epsilon^{\nicefrac{2}{3}}}\left(\frac{2}{3}\right)^{\nicefrac{2}{3}}\frac{\Gamma^2\left(\frac{1}{3}\right)}{\Gamma\left(\frac{11}{6}\right)},\qquad
\Delta\kappa^{(0)}\approx\frac{1}{8}\Delta\kappa^{(\frac{1}{2})},
\end{eqnarray}where  $\Gamma(x)$ denotes  the Gamma function. So, in the limit under consideration the difference between the absorption coefficients decreases
as $\Delta\kappa\sim\xi_\epsilon^{-\nicefrac{2}{3}}$. Meanwhile, the leading order term in   $\Re$ scales as $\sim\xi_\epsilon^{\nicefrac{2}{3}}$,
which turns out to be $\xi_\epsilon^{\nicefrac{4}{3}}$ greater than $\Delta\kappa$. Let us now  turn our attention to  the case  where  $\zeta_\epsilon\ll1$. The latter  is in correspondence with
the conditions $\xi_\epsilon\gg1$ and $\lambda_\epsilon\ll \xi_\epsilon^{-1}$, in which  case  one is  able to  exploit  the large argument behavior of the  Macdonald function, i.e.,
$K_\nu(z)\sim \sqrt{\frac{\pi}{2 z}}e^{-z}$ \cite{Gradshteyn}.  With  this expansion  in  mind,  the variable  $u$ can be  integrated out. As a consequence
\begin{equation}
\Delta\kappa^{(\frac{1}{2})}\approx\frac{\alpha_\epsilon m_\epsilon^2 \lambda_\epsilon}{12\omega}\sqrt{\frac{3}{2}}e^{-\frac{4}{3\zeta_\epsilon}}\qquad  \mathrm{and}\qquad \Delta\kappa^{(0)}\approx\frac{1}{2}\Delta\kappa^{(\frac{1}{2})}.\label{otra}
\end{equation} In this context, we also find  that the leading  expressions of $\Re$ exceed by a factor $\simeq\xi_\epsilon$ the corresponding expression of $\Delta\kappa$.
Clearly, owing to the extinction of the vacuum dichroism the ellipticity   [Eq.~(\ref{rotangle})]  is difficult to  detect. In correspondence,   another kind of observable
is needed to probe the effects induced by the vacuum polarization.

\section{Elastic dispersive properties of the nonlinear vacuum}

The QED  vacuum--polarized by an external field--behaves like a  material medium, in which  light propagation is  modified.  Besides the  dichroic effects,  the  vacuum birefringence is
predicted  to take place: during the  interaction with the strong field of the wave, the   helicity components of the probe beam accumulate a relative difference of the phase. This fact is
closely connected  with the vacuum refraction indices [Eq.~(\ref{refractioninde})]. In correspondence, the incoming linearly polarized probe beam undergoes a tiny rotation [see Fig.~(\ref{fig.000})]
with respect  to the initial polarization plane. This constitutes another observable which is looked for in  the polarimetric  experiments. In the context under consideration, the rotation angle
of the polarization axis  reads
\begin{equation}
\vartheta(\tau)=\frac{1}{2}\frac{n_+-n_-}{n_+n_-}\omega\tau\label{ellipticity}.
\end{equation}  Whenever the dispersive effects are very small, i.e.,  $n_\pm\approx1$,   the denominator of  Eq.~(\ref{ellipticity}) can be taken as unity. The resulting expression resembles   the rotation angle
that  a probe beam undergoes after traversing  a chiral medium.  In the following, we study the regions where Eq.~(\ref{ellipticity}) could be of interest
in the search of MCPs but also in a pure QED context.

\subsection{Photon propagation  at $\xi_\epsilon\lesssim1$ and $\lambda_\epsilon\ll1$}

The kinematic domain where  no absorption of probe photons occurs defines  the transparency region.  Here   the dispersion
relations are  real functions  which  remain   below the  first pair creation threshold, $1<n_*$. However, in the following
we will restrict ourselves to  $1\ll n_*$  where  the  determination of    the   vacuum
refraction indices  [Eq.~(\ref{refractioninde})] is  substantially simplified. To show this,  let us undertake
the calculation of an alternative representation\footnote{A detailed explanation about the operation needed to obtain  Eq.~(\ref{modifiedpi3}) can be found in  Eq.~(27) of Ref.  \cite{VillalbaChavez:2012bb}.\label{footnote}} of $\pi_3^{(0)}$
\begin{eqnarray}\label{modifiedpi3}
\pi_3^{(0)}=-\frac{\alpha_\epsilon m_\epsilon^2\xi_\epsilon^2}{\pi}\int_1^\infty\frac{du}{2u\sqrt{u(u-1)}}\int_{0}^\infty \frac{d\rho}{\rho}e^{-2i u\eta}\left\{\sin^2(\rho) -\frac{8 i}{\lambda_\epsilon} u(u-1)\rho A_0\right\}.
\end{eqnarray} The expression above  involves   $A_0$  [Eq. (\ref{parameters})]  and $\eta$ which is defined below Eq. (\ref{modified}). Whenever
$\xi_\epsilon\lesssim1$ the oscillating term in $\eta$   becomes smaller than the remaining contribution so that
one can approach $\eta\approx \rho/\lambda_\epsilon$. In this limit, the main contribution to the integral over $\rho$ comes  from the region
where  $\rho\sim \lambda_\epsilon\ll1$. Hence,  we can Taylor expand the integrand and obtain
\begin{equation}
 \int_0^{\infty}\frac{d\rho}{\rho}\ldots\simeq \int_0^{\infty}d\rho \ \rho e^{\frac{-2i u \rho}{\lambda_\epsilon}}\left(1-\frac{4iu(u-1)}{3\lambda_\epsilon}\rho\right).
\end{equation} Because of the absence of poles, one can  use  Cauchy's theorem  to rotate  the integration contour by $\rho\to-i\rho$. In correspondence, the variable $\rho$
can be integrated out  and one ends up  with
\begin{eqnarray}\label{modifiedpi3f}
\pi_3^{(0)}\simeq\frac{\alpha_\epsilon m_\epsilon^2\zeta_\epsilon^2}{24\pi}\int_1^\infty\frac{du\ (7-4u)}{u^3\sqrt{u(u-1)}}=\frac{4}{45}\frac{\alpha_\epsilon}{\pi} m_\epsilon^2\zeta_\epsilon^2,
\end{eqnarray}with $\zeta_\epsilon=\lambda_\epsilon\xi_\epsilon$ as before. Besides, following a similar procedure,  we are  able to find  that  $\pi_1^{(0)}\simeq 4 i\alpha_\epsilon m_\epsilon^2\zeta_\epsilon^2\lambda_\epsilon/(105\pi)$,
which  turns out to be  smaller than $\pi_3^{(0)}$ by a factor $\sim \lambda_\epsilon$.  Hence,  for both helicity modes of the probe beam, the  refraction index  [Eq.~(\ref{refractioninde})] is well approached by
\begin{equation}\label{indicesrefractionatximenor1}
n^{(0)}\simeq 1+\frac{2}{45}\frac{\alpha_\epsilon}{\pi}\frac{m_\epsilon^2}{\omega^2}\zeta_\epsilon^2,\qquad\qquad n^{(\frac{1}{2})}\simeq1+\frac{11}{45}\frac{\alpha_\epsilon}{\pi}\frac{m_\epsilon^2}{\omega^2}\zeta_\epsilon^2.
\end{equation} The case  where the  polarization tensor is determined from spin-$\frac{1}{2}$ propagators  has been quoted from Ref.~\cite{baier}.
According to these expressions,   the vacuum in the field of the wave behaves--with  an accuracy up to terms $\sim\lambda_\epsilon$--as a nonbirefringent medium.
Finally, we point out  that in the limits  under consideration  both refraction indices  can be  obtained from the respective  Euler-Heisenberg Lagrangian. Considerations
of this nature have  been carried out in Ref.~\cite{Heinzl,axion} (see also \cite{DiPiazza:2006pr}).

\subsection{Chiral birefringence at $\xi_\epsilon<1$ and $\lambda_\epsilon\gtrsim 1$}

\begin{figure}
\includegraphics[width=3in]{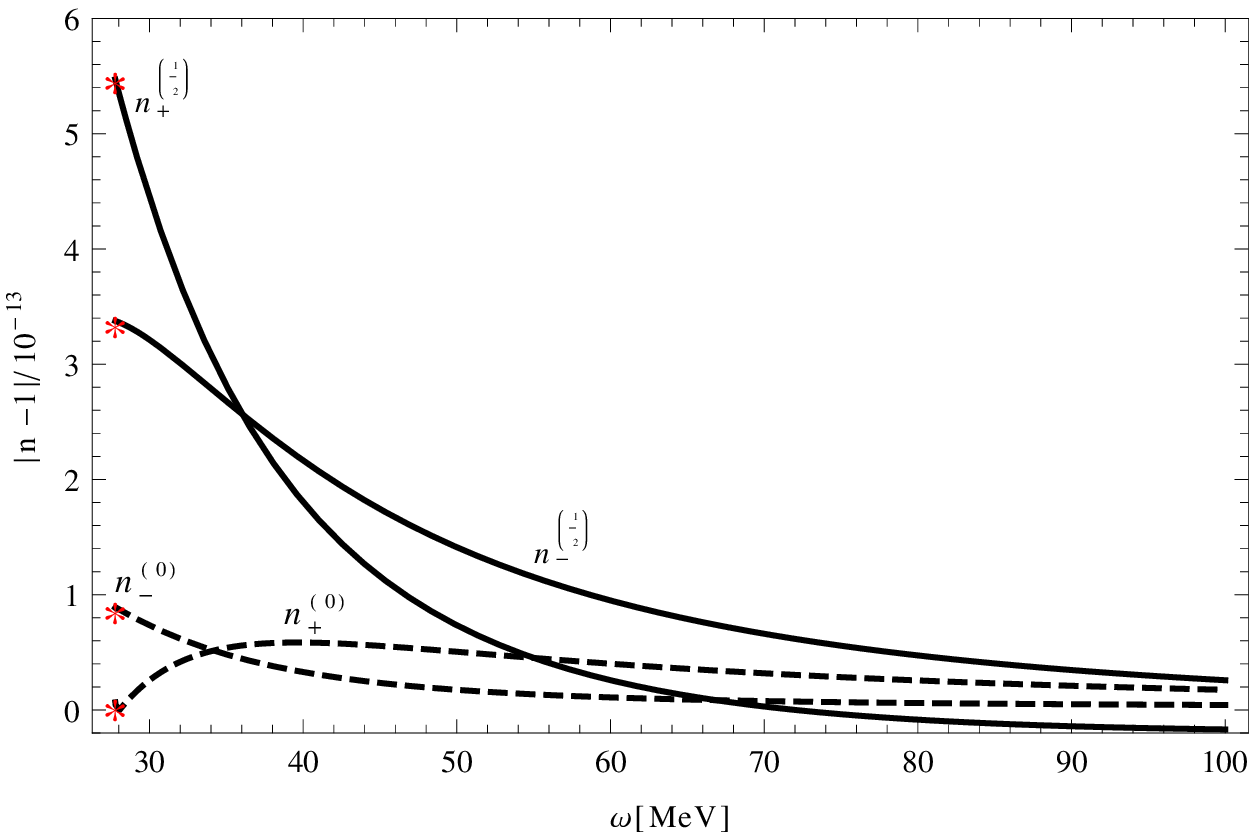}
\includegraphics[width=3.2in]{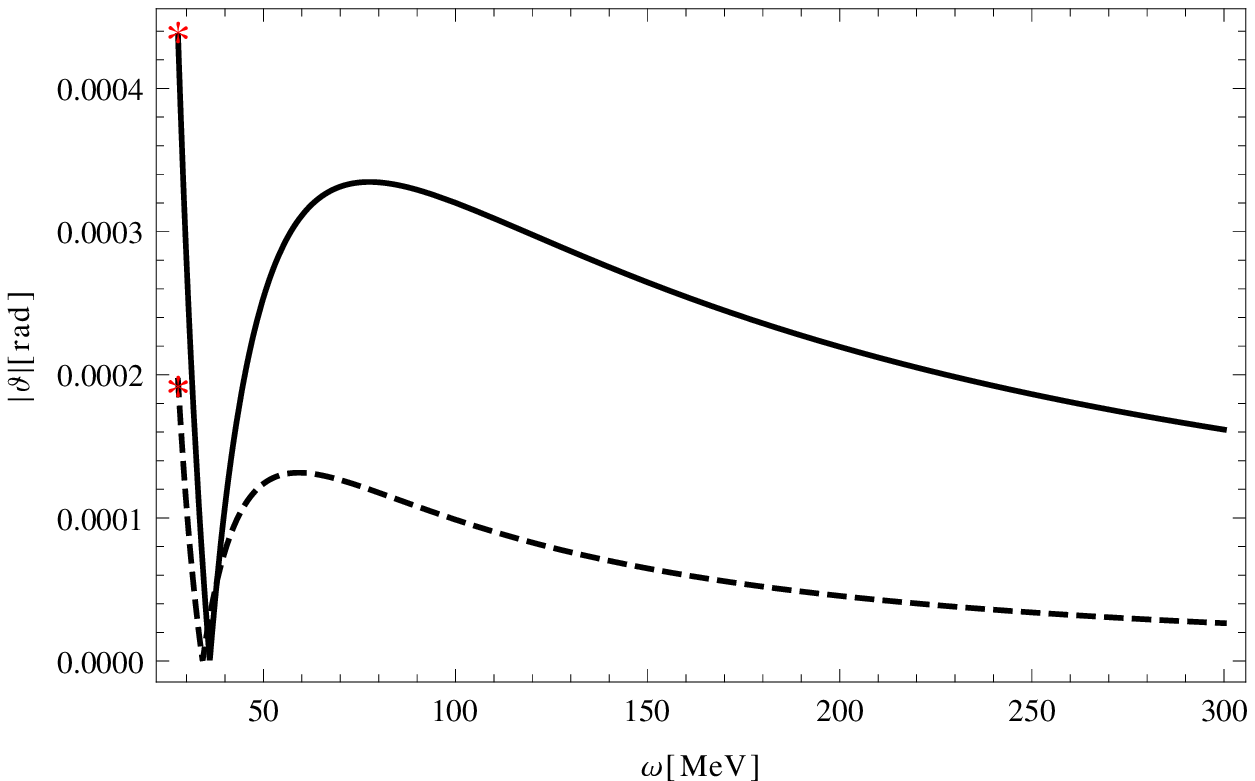}
\caption{\label{fig.005}Correction to the  vacuum  refraction  indices (left) and the rotation of the polarization plane  induced by the vacuum  birefringence  (right) in terms of
the frequency of the probe laser. While the result for spin-$\frac{1}{2}$ particles are represented by  solid curves, those corresponding to the scalar  situation are given by dashed lines.
These results  were obtained by considering  the envisaged XFEL parameters: $\xi=7.5\times 10^{-4}$, $\tau\sim 100\;\text{fs}$,  $\varkappa_0=9\ \rm keV$. The particle charge and
mass equal $e$ and $m$, respectively.}
\end{figure}

The vacuum of virtual $\mathpzc{q}_\epsilon^+\mathpzc{q}_\epsilon^-$ pairs manifests a chiral birefringence when the conditions $\xi_\epsilon<1$ and $n_*\lesssim1$
are simultaneously  fulfilled. To show this, we  start by considering the case where the polarization tensor is determined from spinor  QED. In such a situation, the
real and imaginary parts  of the form  factors involved in the refraction indices [Eq.~(\ref{refractioninde})]
can be approached by
\begin{eqnarray}\label{initialrepi3}
&&\mathrm{Re}\ \pi_3^{(\frac{1}{2})}\simeq\frac{2\alpha_\epsilon m_\epsilon^2\xi_\epsilon^2}{\pi}\ \int_0^1dv\int_{0}^\infty d\rho \frac{\sin^2(\rho)}{\rho} \left\{\frac{n_*}{(1+\xi_\epsilon^2)(1-v^2)\rho}\sin\left(\frac{2n_*\rho}{1-v^2}\right)-\frac{1+v^2}{1-v^2}\cos\left(\frac{2n_*\rho}{1-v^2}\right)\right\},\\
&&\mathrm{Im}\ \pi_1^{(\frac{1}{2})}\simeq\frac{2\alpha_\epsilon m_\epsilon^2\xi_\epsilon^2}{\pi}\ \int_0^1dv\int_{0}^\infty d\rho \frac{1+v^2}{1-v^2}\left\{\frac{\sin^2(\rho)}{\rho^2}-\frac{\sin(2\rho)}{2\rho}\right\}\sin\left(\frac{2n_*\rho}{1-v^2}\right)\label{initialreim1}
\end{eqnarray}where the oscillatory term present in the exponent of Eq.~(\ref{formfactors}) has been neglected. Consequently,  we  will be   working
within the same accuracy limits described in Sec.~\ref{twophotonreaction}. The integral over $\rho$ can, then, be done with  the help of the following identities:
\begin{eqnarray}\label{ss}
\begin{array}{c}\displaystyle
\int_0^\infty\frac{dx}{x^2}\sin^2(x)\sin(2\mathpzc{c}x)=\frac{1}{2}(1+\mathpzc{c})\ln(1+\mathpzc{c})-\mathpzc{c}\ln(\mathpzc{c})-\frac{1}{2}(1-\mathpzc{c})\ln\vert1-\mathpzc{c}\vert \qquad \mathrm{for}\qquad \mathpzc{c}>0,\\ \\
\displaystyle
\int_0^\infty\frac{dx}{x}\sin(2x)\sin(2\mathpzc{c}x)=\frac{1}{4}\ln\left(\frac{1+\mathpzc{c}}{1-\mathpzc{c}}\right)^2 \qquad \mathrm{for}\qquad \mathpzc{c}\neq1, \\ \\ \displaystyle  \int_0^\infty\frac{dx}{x}\sin^2(x)\cos(2\mathpzc{c}x)=\frac{1}{4}\ln\left[\frac{(1+\mathpzc{c})\vert1-\mathpzc{c}\vert}{\mathpzc{c}^2}\right] \qquad \mathrm{for} \qquad \mathpzc{c}>0\qquad \mathrm{and}\qquad \mathpzc{c}\neq1.
\end{array}
 \end{eqnarray} It is worth mentioning at this point that the expression contained in  the first line of Eq.~(\ref{ss}), as well as  the  formula in the
 second line, results  from  an appropriated particularization of  Eqs.~(3.763.3) and (3.741.1) of  Ref.~\cite{Gradshteyn}, respectively.
 The remaining relation is just the derivative of  the expression of the first line.  With these details in mind, one finds that
\begin{eqnarray}\label{interpi3}
&&\mathrm{Re}\ \pi_3^{(\frac{1}{2})}\simeq-\frac{\alpha_\epsilon m_\epsilon^2\xi_\epsilon^2}{\pi}\ \int_0^1dv \left\{-\frac{2n_*}{(1+\xi_\epsilon^2)(1-v^2)}\ln\left[\frac{1-v^2+n_*}{\vert1-v^2-n_*\vert}\right]^{\nicefrac{1}{2}}+\left(\frac{2n_*^2}{(1+\xi_\epsilon^2)(1-v^2)^2}-\frac{1+v^2}{1-v^2}\right)\right.\nonumber\\ &&\qquad\qquad\times\left.\ln\left[\frac{n_*}{\left\vert(1-v^2)^2-n_*^2\right\vert^{\nicefrac{1}{2}}}\right]\right\},\\
&&\mathrm{Im}\ \pi_1^{(\frac{1}{2})}\simeq\frac{\alpha_\epsilon m_\epsilon^2\xi_\epsilon^2}{\pi}\ \int_0^1dv \frac{1+v^2}{1-v^2}\left\{\ln\left[\frac{1-v^2 +n_*}{\vert1-v^2-n_*\vert}\right]^{\nicefrac{1}{2}}-\frac{2n_*}{1-v^2}\ln\left[\frac{n_*}{\left\vert(1-v^2)^2-n_*^2\right\vert^{\nicefrac{1}{2}}}\right]\right\}.\label{interpi1}
\end{eqnarray}The corresponding quantities coming   from  scalar QED  can be read off  from  Eqs.~(\ref{interpi3}) and  (\ref{interpi1}). To do this, one has to  replace  
$(1+v^2)/(1-v^2)\to1$ and   multiply the  right-hand side of these  expressions by a factor $1/2$,  afterwards.  Additionally, the derivation of  $\mathrm{Re}\ \pi_3^{(0)}$  
requires   to change  the signs of the first two terms coming from  Eq.~(\ref{interpi3}). Observe that an exact evaluation 
of the integral over $v$ is quite difficult to perform.  However, when our calculations are particularized with the  QED parameters, i.e., $\epsilon=1$ and $m_\epsilon\to m$, 
it can be  integrated numerically without too much efforts. The resulting corrections to the vacuum refraction indices  are  displayed in Fig.~(\ref{fig.005}). These results 
were obtained by setting the external field parameters to the envisaged XFEL facility.

Now, the dependence  on  $n_*$  allows  us  to obtain--as in  Sec.~\ref{twophotonreaction}--analytical expressions of the vacuum refraction indices. To this end,
we insert Eqs.~(\ref{interpi3}) and (\ref{interpi1}) into Eq.~(\ref{refractioninde}). The presence of the function $\vert1-v^2-n_*\vert$ is then used  to write
the   resulting expression as follows:
\begin{equation}\label{partition}
n_\pm^{(\frac{1}{2})}\simeq 1+\frac{\alpha_\epsilon m_\epsilon^2\xi_\epsilon^2}{2\pi\omega^2}\left\{\int_0^{\sqrt{1-n_*}}dv\left(\ldots\mp\ldots\right)+\int_{\sqrt{1-n_*}}^1dv\left(\ldots\mp\ldots\right)\right\}.
\end{equation}  The respective  integrands  turn out to be   free of  functions involving the absolute value and contain logarithmic divergences at $v=\sqrt{1-n_*}$.
In contrast to the  first one, the last integrand in   Eq.~(\ref{partition})  has an additional divergence at $v=1$. Clearly, whenever $n_*\leq1$,  both  refraction
indices are real, a fact which agrees with the  considerations used in the derivation of Eq.~(\ref{refractioninde}) . Note  that, in  the  region under consideration,
the  photo-production of a $\mathpzc{q}_\epsilon^+\mathpzc{q}_\epsilon^--$pair
could take place
[see Sec.~\ref{twophotonreaction}].  Therefore,  the refraction indices in Eq.~(\ref{partition}) describe  the dispersive properties of   those  photons that--having
the  proper energies-- do not take part in  the two-photon reaction.
It is precisely in a vicinity of the corresponding threshold  [$n_*\simeq1$] where the  chiral birefringence effect turns out to be  maximized.  This is
manifest within the QED context [see Fig.~(\ref{fig.005})]. Hence, finding  expressions which describe the situation in this particular limit is also of
interest. To this end, we set $n_*=1$ and compute the relevant  integral by using MATHEMATICA code. As a consequence,
\begin{eqnarray}\label{puntual}
\left.n_\pm^{(\frac{1}{2})}\right\vert_{n_*=1}\approx1+\frac{\alpha_\epsilon m_\epsilon^2\xi_\epsilon^2}{2\pi\omega^2}\left(0.9+\frac{1.2}{1+\xi^2_\epsilon}\pm0.5\right), \qquad\left.n_\pm^{(0)}\right\vert_{n_*=1}\approx1+\frac{\alpha_\epsilon m_\epsilon^2\xi_\epsilon^2}{4\pi\omega^2}\left(0.8-\frac{1.2}{1+\xi^2_\epsilon}\pm0.4\right)
\end{eqnarray}where the outcome resulting  from scalar QED has been included. The explicit expression of the rotation angle follows  by  inserting Eq.~(\ref{puntual})
into Eq.~(\ref{rotangle}). Consequently,
\begin{eqnarray}\label{rotanglen1}
\left.\vartheta^{(\frac{1}{2})}(\tau)\right\vert_{n_*=1}\approx\frac{\alpha_\epsilon  m_\epsilon^2 }{4\pi\omega} \xi_\epsilon^2\tau\qquad \mathrm{and}\qquad\left.\vartheta^{(0)}(\tau)\right\vert_{n_*=1}\approx0.4\left.\vartheta^{(\frac{1}{2})}(\tau)\right\vert_{n_*=1}.
\end{eqnarray} We remark that, in both cases,  the rotation angle enhances when  the
frequency of the  probe beam is small  and the product $\xi_\epsilon^2\tau$ becomes  large. It is also convenient to emphasize that Eq.~(\ref{rotanglen1}) is applicable
only when  $m_\epsilon=[(\varkappa k)/2]^{\nicefrac{1}{2}}$. 
As last remark of  this subsection,  we  point  out  that the expressions associated with the first Born [$\xi_\epsilon\ll1$] approximation can be obtained from Eq.~(\ref{puntual})
 just  by setting  $\xi_\epsilon^2=0$ in  the fraction contained  within  the brackets.

\subsection{Large asymptotic behavior of the vacuum refraction indices  at $\xi_\epsilon\gg1$ \label{dipersivestrongfield}}

Our aim in this subsection is to determine   the asymptotic behavior of those  quantities associated  with the vacuum birefringence as  $\xi_\epsilon\gg1$.
The problem under consideration is quite analogous to  the one analyzed in Sec.~\ref{quasi}, so that the line of reasoning  will be similar. However, some  differences will emerge in the course
of the calculations. These differences come out   from the nature
of $\mathrm{Re}\  \pi_1$ and $\mathrm{Im}\ \pi_3$ involved in Eq.~(\ref{refractioninde}). Let us undertake,  in first instance, the computation of the difference
between the vacuum refraction indices. Within the framework of spinor QED, and with the  help of Eqs.~(\ref{refractioninde}) and  (\ref{formfactors})-(\ref{parameters}),
the latter turns out to be
\begin{equation}\label{deltan}
\Delta n^{(\frac{1}{2})}(\lambda_\epsilon,\xi_\epsilon)=n_+^{(\frac{1}{2})}-n_-^{(\frac{1}{2})}\simeq-\frac{\mathrm{Im}\ \pi_1^{(\frac{1}{2})}}{\omega^2}\simeq-\frac{4\alpha_\epsilon m_\epsilon^2\xi_\epsilon^2}{\pi\omega^2}\int_1^\infty\frac{du (2u-1)}{2u\sqrt{u(u-1)}}\int_0^\infty  d\rho A_0\sin\left(2u\eta\right).
\end{equation}We follow  the premise of dividing the $\rho-$interval  into two domains:  from  $0$ to $\rho_0$ and from $\rho_0$ to $\infty$ with $\xi_\epsilon^{-1}\ll\rho_0\ll1$.
Once this step has been  carried out,   we are allowed to  Taylor expand $A_0\approx\rho^2/6$ and  $\eta=\frac{\rho}{\lambda_\epsilon}\left[1+\xi_\epsilon A\right]\approx\frac{1}{\zeta_\epsilon}\left[\rho\xi_\epsilon+\frac{(\rho\xi_\epsilon)^3}{3}\right]$
separately in the interval where  $\rho\in[0,\rho_0]$.  In correspondence,  Eq.~(\ref{deltan}) approaches to
\begin{equation}\label{firstapproach}
\Delta n^{(\frac{1}{2})}\simeq-\frac{\alpha_\epsilon m_\epsilon^2}{3 \pi\omega^2 \xi_\epsilon}\int_1^\infty\frac{du (2u-1)}{u\sqrt{u(u-1)}}\int_0^\infty ds s^2\sin\left[\frac{2u}{\zeta_\epsilon}\left(s+\frac{s^3}{3}\right)\right].
\end{equation}The derivation of this equation requires to perform the  change of variable $\rho\xi_\epsilon\to s$  and to  extend  the resulting integration limit to infinity
[$\rho_0\xi_\epsilon\to\infty$].  We remark that Eq.~(\ref{firstapproach})  turns out to be a good approximation only if the exponential $\sim\exp\left[-\frac{2u}{3\zeta_\epsilon}(\rho_0\xi_\epsilon)^3\right]$  
falls off  sufficiently fast as
$\rho_0\xi_\epsilon\gg1$. Of course, the latter condition is satisfied whenever the exponent is very large,  in which case  the  complementary restriction [Eq.~(\ref{constraint})]
$1\gg \rho_0\gg (\lambda_\epsilon/\xi_\epsilon^2)^{\nicefrac{1}{3}}$ is found. Formally, it  should appear an additional dependence  on the variable $u$. Nonetheless,  we have set $u\sim 1$ because
 the vicinity of this value  provides the main contribution of the $u-$integral.

Now, we  perform the change of variable $(2u/\zeta_\epsilon)^{\nicefrac{1}{3}}s\to y$ and  express
the integral over $s$ in terms of the second derivative of  Scorer's function \cite{Olver}\footnote{The name of this function varies in the literature. A summary of its properties
can be found  on page 448 of Ref. \cite{abramowitz}. Another compact recap is  given in Appendix E of Ref. \cite{Katkov} with the name of Hardy's function $\Upsilon(z)$. In \cite{ritus}, the name Upsilon function is used. A relation between both notations can be established according to $\Upsilon(z)=\pi\mathrm{Gi}(z)$.}:
\begin{equation}
\int_0^\infty ds\ldots=\frac{\zeta_\epsilon \pi}{2u}\mathrm{Gi}^{\prime\prime}\left(z\right)\qquad \mathrm{with}\qquad
\mathrm{Gi}(z)=\frac{1}{\pi}\int_0^\infty dy \sin\left(z y+\frac{y^3}{3}\right)\qquad \mathrm{and}\qquad z\equiv \left(\frac{2u}{\zeta_\epsilon}\right)^{\nicefrac{2}{3}}. \label{scorer}
\end{equation}We insert this expression into Eq.~(\ref{firstapproach}) and  use  the differential equation $ \mathrm{Gi}^{\prime\prime}\left(z\right)-z\mathrm{Gi}(z)=-\pi^{-1}$
afterwards. With these steps in mind,  the differences between the vacuum refraction indices turn out to be
\begin{eqnarray}\label{firstapproachvariante1}
&&\Delta n^{(\frac{1}{2})}\simeq-\frac{4\alpha_\epsilon m_\epsilon^2\lambda_\epsilon}{9 \pi\omega^2}+\frac{2^{\nicefrac{2}{3}}\alpha_\epsilon m_\epsilon^2\lambda_\epsilon}{6 \omega^2\zeta_\epsilon^{\nicefrac{2}{3}}}\int_1^\infty\frac{du (2u-1)}{u^{\nicefrac{4}{3}}\sqrt{u(u-1)}}\mathrm{Gi}(z),\\
&&\Delta n^{(0)}\simeq-\frac{2\alpha_\epsilon m_\epsilon^2\lambda_\epsilon}{9 \pi\omega^2}+\frac{2^{\nicefrac{2}{3}}\alpha_\epsilon m_\epsilon^2\lambda_\epsilon}{12 \omega^2\zeta_\epsilon^{\nicefrac{2}{3}}}\int_1^\infty\frac{du}{u^{\nicefrac{4}{3}}\sqrt{u(u-1)}}\mathrm{Gi}(z).\label{firstapproachvariante2}
\end{eqnarray} Note   that Eqs.~(\ref{firstapproachvariante1}) and (\ref{firstapproachvariante2}) depend--as for $\Delta\kappa$ in Sec.~\ref{quasi}--on the frequency
of the high-intensity laser, a fact which   does not find a counterpart in the constant crossed field approach.

It is interesting to  proceed by  restricting  the parameter $\zeta_\epsilon$ to some asymptotic limits of interest. We start with  the situation in which  $\zeta_\epsilon\gg1$ [corresponding to
$\lambda_\epsilon\gg 1/\xi_\epsilon$ with $\xi_\epsilon\gg1$].  Considering  the appropriate  expansion  of  Scorer's function at $z\ll1$, i.e.,
$\mathrm{Gi}(z)\sim\frac{1}{ 2\pi\  3^{\nicefrac{2}{3}}}\Gamma\left(\frac{1}{3}\right)+\frac{1}{2\pi\ 3^{\nicefrac{1}{3}}}\Gamma\left(\frac{2}{3}\right) z$, one  obtains
\begin{eqnarray}
\Delta n^{(\frac{1}{2})}\simeq\frac{4 \alpha_\epsilon m_\epsilon^2 \lambda_\epsilon}{9\pi\omega^2}+ \frac{\sqrt{3}}{3\omega}\Delta\kappa^{(\frac{1}{2})}\quad\mathrm{and}\qquad\Delta n^{(0)}\simeq\frac{2 \alpha_\epsilon m_\epsilon^2 \lambda_\epsilon}{9\pi\omega^2}+ \frac{\sqrt{3}}{2\omega}\Delta\kappa^{(0)}.\label{smallzetan}
\end{eqnarray}We point out that the quantity $\Delta\kappa$ can be found in Eq.~(\ref{largekappazeta}). Since it is suppressed by a factor $\sim1/\xi_\epsilon$ one can ignore
its contribution and just  deal  with the leading order terms.  The latter are  independent of  the parameter $\xi_\epsilon$  and  maximized when the collision between  the
probe and  the external   wave is head-on.  We should also mention that, although the leading term  is  independent of the mass of the particle, it  applies  for those values
with  $m_\epsilon\ll[\epsilon (k\varkappa) m/2]^{\nicefrac{1}{3}}$. Moreover, the rotation angle, which comes out  of combining the
expression for $\Delta n^{(\frac{1}{2})}$  with Eq.~(\ref{ellipticity}), is independent of the frequency of the probe beam. Considering the configuration in which both lasers
counterpropagate,  we find
 \begin{equation}
 \vartheta^{(\frac{1}{2})}(\tau)=\frac{2\alpha }{9\pi}\epsilon^2\varkappa_0\tau\label{strongellipticity}
\end{equation}where  $\tau$ is  the interacting time  and $\alpha=1/137$  the QED fine structure constant.  Observe that a comparison with the rotation angle
coming out from the scalar case leads to write  $\vartheta^{(\frac{1}{2})}\approx2\vartheta^{(0)}$.

The situation is slightly  different in  the case where $\zeta_\epsilon\ll1$. This condition restricts  $\lambda_\epsilon\ll1$ with  $\xi_\epsilon\gg1$. In this context, the large
asymptotic behavior of  Scorer's  function applies, i.e.,  $\mathrm{Gi}(z)\sim \frac{1}{\pi z}+\frac{2}{\pi z^4}$.  Consequently, we can develop the   integral    over $u$
and find that
\begin{equation}\label{largezeta}
\Delta n^{(\frac{1}{2})}\simeq\frac{32\alpha_\epsilon}{315\pi}\frac{m_\epsilon^2}{\omega^2}\frac{\zeta_\epsilon^2 }{\xi_\epsilon}\qquad\mathrm{and}\qquad\Delta n^{(0)}\simeq \frac{n^{(\frac{1}{2})}}{4}.
\end{equation}Accordingly,  a suppression $\sim \xi_\epsilon^{-1}$ of $\Delta n$ occurs. Therefore, under the aforementioned  circumstance,  the nonlinear vacuum of QED seems to  behave
as a  material in which dichroism [Eqs.~(\ref{largekappazeta}) and (\ref{otra})] and birefringence are practically absent. We will shortly retake this point again.

We want to conclude this section by  determining the  expression of the vacuum refraction indices. According to Eq.~(\ref{refractioninde}) and (\ref{deltan}),  it can be
written as
\begin{equation}
n_\pm^2-1=\frac{\mathrm{Re}\ \pi_3}{\omega^2}\pm\Delta n.\label{varianterefractionidex}
\end{equation}
We have already determined the leading behavior of $\Delta n$ as $\xi_\epsilon\gg1$. So, our goal now is to compute the
isotropic contribution $\sim\rm Re\ \pi_3/\omega^2$. To undertake the calculation,  we first integrate by parts those  terms  of the integrand  $\Omega_3$  [Eq.~(\ref{g3})] proportional
to $\sim(1-e^y)$. This step  allows us to express the contribution
resulting from spinor QED in the following form:  (see footnote \ref{footnote})
\begin{equation}
\frac{\mathrm{Re}\ \pi_3^{(\frac{1}{2})}}{\omega^2}=-\frac{2\alpha_\epsilon m_\epsilon^2\xi_\epsilon^2}{\pi\omega^2}\int_1^\infty\frac{du}{2u\sqrt{u(u-1)}}\int_0^\infty\frac{d\rho}{\rho}\left\{(2u-1)\sin^2(\rho)\cos(2u\eta)+\frac{8u(u-1)}{\lambda_\epsilon}\rho A_0\sin(2u\eta)\right\}\label{intermn}.
\end{equation}What remains is to apply the preceding method to Eq.~(\ref{intermn}). Carrying out the appropriate steps  and with the  help of the representation of  Scorer's
function [Eq.~(\ref{scorer})] we end up with
\begin{equation}
\frac{\mathrm{Re}\ \pi_3^{(\frac{1}{2})}}{\omega^2}\simeq-\frac{\alpha_\epsilon m_\epsilon^2\zeta_\epsilon^{\nicefrac{2}{3}}}{3\omega^2 2^{\nicefrac{2}{3}}}\int_1^\infty\frac{du\ (8u+1)}{2u^{\nicefrac{5}{3}}\sqrt{u(u-1)}}\mathrm{Gi}^\prime(z),
\qquad \frac{\mathrm{Re}\ \pi_3^{(0)}}{\omega^2}\simeq-\frac{\alpha_\epsilon m_\epsilon^2\zeta_\epsilon^{\nicefrac{2}{3}}}{6\omega^2  2^{\nicefrac{2}{3}}}\int_1^\infty\frac{du\ (4u-1)}{2u^{\nicefrac{5}{3}}\sqrt{u(u-1)}}\mathrm{Gi}^\prime(z)\label{realpi3}
\end{equation} where the expression  resulting  from scalar QED has been included. We  then consider the case $\zeta_\epsilon\gg1$ and  insert the small argument behavior of
the $\mathrm{Gi}(z)$ into Eq.~(\ref{realpi3}).  The obtained  expression exceeds  the birefringent term  [Eq.~(\ref{smallzetan})] by a factor $\sim (\xi_\epsilon^2/\lambda_\epsilon)^{\nicefrac{1}{3}}$.
According to the complementary condition [see below Eq.~(\ref{firstapproach})], this  factor must be  much greater than unity.  In correspondence, the vacuum refraction index
for both helicity modes of the probe beam approaches to
\begin{eqnarray}
n^{(\frac{1}{2})}\approx1-\frac{5\alpha_\epsilon m_\epsilon^2 \zeta_\epsilon^{\nicefrac{2}{3}}}{36 \sqrt{\pi}\omega^2}\left(\frac{3}{2}\right)^{\nicefrac{2}{3}}\frac{\Gamma^2\left(\frac{2}{3}\right)}{\Gamma\left(\frac{13}{6}\right)} 
\qquad
n^{(0)}\approx1-\frac{\alpha_\epsilon m_\epsilon^2 \zeta_\epsilon^{\nicefrac{2}{3}}}{36\sqrt{\pi}\omega^2}\left(\frac{3}{2}\right)^{\nicefrac{2}{3}}\frac{\Gamma^2\left(\frac{2}{3}\right)}{\Gamma\left(\frac{13}{6}\right)}.
\end{eqnarray}   The latter result means that the vacuum in the field of a circular polarized  wave, in which  $\zeta_\epsilon\gg1$, behaves like   a quasi-nonbirefringent
crystal where the rotation [Eq.~(\ref{strongellipticity})] comes from a tiny birefringent effect.

\begin{wrapfigure}{r}{0.5\textwidth}
\includegraphics[width=.5\textwidth]{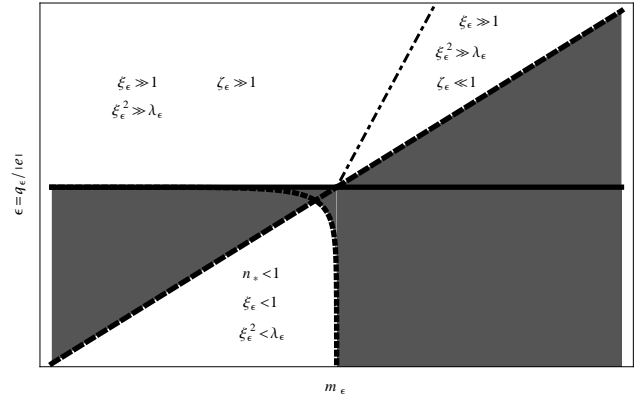}
\caption{\label{fig.002} Regions of applicability of our calculations in the search  of MCPs  are indicated in the upper and lower  white sectors. While  the dashed line corresponds
to   $\epsilon m \xi/m_\epsilon=1$,  the solid line   represents $\epsilon^2\xi^2=\lambda$.  The dotted  curve  comes out  from the condition associated with the  first pair creation 
threshold  $n_*=1$. The dash-dotted line  arises from the equation $\zeta_\epsilon=1$.  The gray shadow regions  cannot be explored  with the  approximations used in this work.}
\end{wrapfigure}

This situation  is even more pronounced  when the opposite condition  $\zeta_\epsilon\ll1$ is taken into account. Here the  photon energy $\omega$ lies below the first pair creation threshold
region where  $\lambda_\epsilon\ll\xi_\epsilon^{-1}\ll1$ but $z\gg1$  and in correspondence one   can use the large asymptotic behavior of $\mathrm{Gi}(z)$ [given  above  Eq.~(\ref{largezeta})]
in Eq.~(\ref{realpi3})  to find that
the leading order correction  of the vacuum refractive index resembles  the one arising   in the limit of $\xi_\epsilon\lesssim1$ [Eq.~(\ref{indicesrefractionatximenor1})].  Thus, this result ratifies  our previous claim about the nonbirefringent
character of the vacuum when  $\zeta_\epsilon\ll1$.

We therefore find  that, in connection with the dying out of the dichroic phenomenon at $\xi_\epsilon\gg1$,  an extinction of   the  vacuum birefringence takes place as  well.
In such a case,  it would be convenient to have another observable  at  our disposal which helps us to investigate the effects induced by the MCPs [see Secs. \ref{ramaninelastwave1}
and \ref{rammanwavespoltensor} below]. However,
before  we will show  that  the region around the first pair creation threshold,  where the dichroism and birefringence are manifest, provides interesting bounds.

\section{Laser-assisted search of MCPs}

\subsection{Perspectives  in the  birefringent and dichroic sector \label{perspective}}

The prospect of finding exclusion limits  on the MCPs using laser technology is certainly enticing. In polarimetry, the idea is to use Eq.~\eqref{rotangle}
or   Eq.~(\ref{ellipticity}) to restrict the parametric space defined by the  $(\epsilon,m_\epsilon)$ plane  in the way that a  high-precision optical
measurement of   $\psi(\tau)$ or $\vartheta(\tau)$ is carried out without a significant  detection of the effects induced by the MCPs. This requirement
implies, for instance,  that the  sensitivity level in the experiment $\psi_{95\% CL}$--which we suppose verified at $95\%$ confidence level (CL)--is not
high enough  for observing  the hypothetical ellipticity   due to the photo-production of a $\mathpzc{q}_\epsilon^+\mathpzc{q}_\epsilon^-$pair. A similar idea
applies when a measurement of the rotation of the polarization plane  is carried out without  success. As a consequence, the  relations $\psi_{95\% CL}>\psi(\tau)$
and $\vartheta_{95\% CL}>\vartheta(\tau)$  [with  Eqs.~\eqref{rotangle} and \eqref{ellipticity}  included] must be understood  as the starting point  for
finding  out  the constraints on the  parameters associated with the  minicharged
carriers.

We start our analysis by considering   the regions   where the dichroism and birefringence of the vacuum are strongly manifest, i.e.,   where $\xi_\epsilon<1$
and $n_*\lesssim1$. The results coming out from this regime  can be expected to be trustworthy when the bound  is embedded between  the  dashed line, corresponding
to  $\epsilon m \xi/m_\epsilon=1$ and the  dotted line corresponding to $n_*=1$.  This region of applicability is displayed in Fig.~\ref{fig.002}  [lower white sector],
which  must be understood in a log-log scale. Note that the shape of this figure is generic, it does not depend on the special value of $\xi$ chosen. Once the parameters of
the strong wave are fixed, the region encompassed between the aforementioned curves  cannot be studied with the approximations used in this work. Obviously,
the  bounds to be found in this kind of high precision optical experiment  depend primarily  on the intrinsic properties of both laser beams.
Regarding  this point we  precise  that,  since  our  external field [Eq.~(\ref{externalF})] is a monochromatic plane-wave,  an appropriated  experimental
setup should incorporate an  intense source where its   oscillating period is much smaller than its   temporal extension [$\tau\gg 2\pi \varkappa_0^{-1}$].

So, when evaluating  our expressions  we have in mind an  achievable  experimental  condition in which $\xi\approx 6.5 \times10^{-2}$, $\tau\approx 20\ \rm ns$ and $\varkappa_0\simeq 1.2 \ \rm eV$.
Such parameters correspond to  the Petawatt High-Energy Laser for heavy Ion eXperiments (PHELIX)  \cite{PHELIX}, currently under operation in Darmstadt, Germany. For the probe beam we will choose
an optical laser with $\omega=2\varkappa_0\simeq 2.4\ \rm eV$.  In principle, this may be obtained by frequency-doubling of a portion extracted from  the strong wave.
Let us suppose a polarimetric experiment where the rotation [Eq.~(53)] is probed. Assuming a slightly sub-resonant mass  $m_\epsilon\approx1.7\ \rm eV$ [corresponding to $n_*\approx1$], the  constraint
resulting for  spinor MCPs is  $\epsilon< 1.9 \times 10^{-6}$. On the contrary, for Klein-Gordon particles we find that $\epsilon<2.3\times10^{-6}$ applies.  These  results have been determined
by supposing the counterpropagating geometry and by considering   sensitivities  of the order of $\sim 10^{-10}\ \rm rad$, which appears  achievable \cite{Muroo_2003}.  Note that for
$m_\epsilon\sim 1\ \mathrm{eV}$ and the relevant range of $\epsilon$, the chosen intensity parameter $\xi$ corresponds to $\xi_\epsilon=\xi\frac{m}{m_\epsilon} \epsilon\ll1$.  It is
worth  emphasizing at this point  that our   predictions  cover regions  of  masses in which the constraints--deduced from several experimental  collaborations--are less restrictive
\cite{Ahlers:2007qf,Ahlers:2006iz}.  Therefore, high-precision  optical experiments  in  a laser wave of moderate intensities  [$\xi_\epsilon\leqslant1$]  can complement  the MCP
searches at  dipole magnets.

Less stringent constraints are  found in the strong field regime $\xi_\epsilon\gg 1$ when  the rotation of the polarization plane  [Eq.~(\ref{strongellipticity})]
is taken into account. The  bounds which  arise  by using this expression  must  be consistent with  the  conditions $\xi_\epsilon\gg1$, $\xi_\epsilon^2\gg\lambda_\epsilon$ and $\lambda_\epsilon\xi_\epsilon\gg1$ under which it was derived. Correspondingly, they have to be located far to the left of the
curve  $\zeta_\epsilon=\epsilon m^3 \lambda\xi/m_\epsilon^3=1$ but also high above  the curve    $\epsilon^2\xi^2=\lambda$. This region can be seen in Fig.~\ref{fig.002}
as well [upper white sector]. We then particularize Eq.~(\ref{strongellipticity}) with the parameters associated with  the Petawatt Optical Laser Amplifier for Radiation Intensive
experimentS (POLARIS) \cite{POLARIS}, presently  under operation  in Jena, Germany.  We remark that the intensity parameter related to this laser system is expected to reach an order of magnitude
$\xi\sim 10^2$ which justifies  its use in the strong  field approach. This would most probably be acheived by compressing  $120\ \rm J$ at pulse lengths $\tau_{\mathrm{p}}=120\ \rm fs$, i.e. a power near
$1\ \rm PW$. Moreover, this  laser   operates  with a central wavelength $\lambda_0=1035\ \rm nm$ corresponding   to  a  frequency $\varkappa_0=1.2\ \rm eV$,
whose combination  with  its temporal length  guarantees the monochromaticity  condition $\varkappa_0\tau_{\mathrm{p}}\gg1$.  When the probe beam is chosen as an optical
laser too--similar to the Multi-Terawatt class laser JETI--[$\omega=1.55\ \rm eV$, $\tau_{\mathrm{J}}=30\ \rm fs$] a sensitivity level  $\vartheta\sim10^{-10}\ \rm rad$ seems
to be reasonable \cite{Muroo_2003}. By choosing $\tau=\tau_{\mathrm{J}}$, we found,
for Dirac Fermions  $\epsilon<6.0\times10^{-5}$. In contrast,  for Klein-Gordon particles $\epsilon<8.5\times10^{-5}$. We remark that  both results apply for
masses below the  eV regime.

\subsection{Inelastic scattered waves: the quasi-monochromatic approach \label{ramaninelastwave1}}

The suppression of the  vacuum dichroism  in  the  strong field regime  $\xi_\epsilon\gg 1$, together with the weakness of its  birefringence property,
motivate us  to look for  an observable different from  the ellipticity and the rotation of the polarization plane. Among the plausible options, we choose
for counting  the number of Raman-like  photons  which are  generated from the inelastic interaction, i.e., the last two terms in  Eq.~(\ref{ptcircular}).
In this context,  the  rate of detected photons  whose momentum  differs  from the incoming
probe beam  is given by
\begin{equation}
\dot{\mathpzc{N}}=\dot{\mathpzc{N}}_{0}\mathpzc{P}_{\gamma\to\gamma^\prime}
\end{equation} where  an optimal efficiency of detection has been assumed. Here $\dot{\mathpzc{N}}_{0}$ denotes the number of incoming photons per unit
of time and $\mathpzc{P}_{\gamma\to\gamma^\prime}$ is the  respective generation probability.  The latter  can be computed from  Eq.~(\ref{ptcircular}) since,
on the mass-shell [$k^2=0$], the polarization tensor defines the photon-photon scattering amplitude with  a  potential change  of polarization $\mathpzc{e}_{1}^\mu\to \mathpzc{e}_{2}^\mu$,
i.e., $\mathrm{T}_{\mathpzc{e}_2k_2,\mathpzc{e}_1k_1}=\mathpzc{e}_2^\mu\Pi_{\mu\nu}(k_2,k_1)\mathpzc{e}^\nu_1/[2V(\omega_2 \omega_1)^{\nicefrac{1}{2}}]$
with  the volume $V$ where the interaction takes place.  The  generation  of  Raman-like electromagnetic waves occurs whenever  $\mathpzc{e}_{1}=\mathpzc{e}_{2}=\Lambda_\pm/2^{\nicefrac{1}{2}}$.
In correspondence, the total production rate reads
\begin{eqnarray}\label{ratefermionphotongeneration}
\mathpzc{R}_{\gamma\to\gamma^\prime}=\frac{\vert\pi_0(k+2\varkappa)\vert^2}{\omega_{\pmb{k}+2\pmb{\varkappa}}\omega_{\pmb{k}}}2\pi\delta(\omega_{\pmb{k}}-\omega_{\pmb{k}+2\pmb{\varkappa}}+2\varkappa_0)
+\frac{\vert\pi_0(k-2\varkappa)\vert^2}{\omega_{\pmb{k}-2\pmb{\varkappa}}\omega_{\pmb{k}}}2\pi\delta(\omega_{\pmb{k}}-\omega_{\pmb{k}-2\pmb{\varkappa}}-2\varkappa_0)
\end{eqnarray}where  $\omega_{\pmb{k}\pm2\pmb{\varkappa}}\equiv\vert\pmb{k}\pm2\pmb{\varkappa}\vert$ denotes the frequencies of the Raman-like waves.
Eq.~(\ref{ratefermionphotongeneration})  does not contain interference terms since the product of delta functions  with different momentum contents  vanishes
identically. The energetic   balances imposed by the Dirac deltas in Eq.~(\ref{ratefermionphotongeneration}) cannot be fulfilled though  unless $\pmb{k}$ and
$\pmb{\varkappa}$ are collinear, in which case the form factors are zero. As a consequence,
the  respective   rate $\mathpzc{R}_{\gamma\to\gamma^\prime}$ vanishes identically always.
The appearance of these Dirac's deltas is intrinsically connected with
the monochromaticity   of the strong field of the wave.   They  are distorted to another distribution functions when  a finite laser  pulse is taken into
account. This is, in fact, the case in which we are interested in. So, in the following it must be understood that our infinite plane wave train is  an
approximation to the situation of practical interest, where the product of $\varkappa_0\tau$ is very large but, on the other hand, of a finite value.

Now,  the   form factor involved in Eq.~(\ref{ratefermionphotongeneration})  is given by
\begin{equation}\label{pio0}
\pi_0^{(\frac{1}{2})}(k\pm2\varkappa)=-\frac{\alpha_\epsilon}{\pi} m_\epsilon^2\xi_\epsilon^2\int_{0}^{1}dv \int_0^\infty \frac{d\rho}{\rho}e^{-\frac{2i\rho}{\vert\lambda_\epsilon\vert(1-v^2)}\left[1+2A\xi_\epsilon^2\right] \pm 2i\rho \mathrm{sign}[\lambda_\epsilon]}A_1
\end{equation}where  the quantities involved in this formula can be found in Eqs.~(\ref{vectorialbasisbaier})-(\ref{parameters}).  Hereafter  we   focus ourselves
to the case where  the parameter  $\xi_\epsilon\gg1$. In this context, the  large asymptotic behavior of $\pi_0$  follows from the application of the method previously
implemented in Secs.~\ref{quasi} and \ref{dipersivestrongfield}. The only difference stems  in an additional factor proportional to  $\mathrm{sign}[\lambda_\epsilon]$,
present in the exponent of Eq.~(\ref{pio0}). In correspondence, one finds that
\begin{equation}\label{cuucucuuu}
\pi_0^{(\frac{1}{2})}(k\pm2\varkappa)\simeq-\frac{\alpha_\epsilon m_\epsilon^2\zeta_\epsilon^{\nicefrac{2}{3}}}{2^{\nicefrac{5}{3}}}\int_1^\infty\frac{du}{2u^{\nicefrac{5}{3}}\sqrt{u(u-1)}}\left[\mathrm{Gi}^\prime(z_\pm)-\frac{i}{\sqrt{3}}z_\pm K_{\nicefrac{2}{3}}\left(\frac{2}{3}z_\pm^{\nicefrac{3}{2}}\right)\right]
\end{equation}where the arguments of the special functions $\rm Gi$ and $K_{\nicefrac{2}{3}}$ read $z_\pm=(2u/\zeta_\epsilon)^{\nicefrac{2}{3}}\left[1\mp\lambda_\epsilon/u \right]$.
As the expressions derived in Secs.~\ref{quasi} and \ref{dipersivestrongfield}, Eq.~(\ref{cuucucuuu}) applies whenever  the number of absorbed photons is very large
with $\xi_\epsilon^2\gg\lambda_\epsilon$. Therefore, the  bounds that  emerge by using this formula must  be located far to the left of the curve  $\epsilon m \xi/m_\epsilon=1$
but also far  above from   $\epsilon^2\xi^2=\lambda$  [see Fig.~\ref{fig.002}]. Next, for asymptotically large value of  $\zeta_\epsilon\gg1$  the  arguments $z_\pm\sim0$. As
long as  the  expansions in the small argument of $\mathrm{Gi}$ and $K_{\nicefrac{2}{3}}$  are  used, the expression above acquires the  simpler structure
\begin{equation}\label{imprt}
\pi_0^{(\frac{1}{2})}(k\pm2\varkappa)\approx-\frac{\alpha_\epsilon m_\epsilon^2\zeta_\epsilon^{\nicefrac{2}{3}}\Gamma^2\left(\frac{2}{3}\right)}{42 \sqrt{\pi}\ \Gamma\left(\frac{7}{6}\right)}\left(\frac{2}{3}\right)^{\nicefrac{2}{3}}\left(1-i\sqrt{3}\right).
\end{equation}As an outcome of the previous analysis,  we observe that  in the limit under consideration, $\pi_0^{(\frac{1}{2})}(k\pm2\varkappa)\approx\pi_0^{(\frac{1}{2})}(k)$.

In order to evaluate the effects coming   from the finite size of the strong laser beam,  we   determine the explicit  solution of the equation
of motion [Eq.~(\ref{sdpmBF})] and  impose  boundary conditions afterwards. A substantial simplification of the problem  is achieved by ignoring the
contributions given by $\pi_1$ and $\pi_3$ in   Eq.~(\ref{dispr1})  and  keeping only the off-diagonal form factors.  The diagonal quantities  are then
linearized  according to the rules $k^2\simeq 2\omega_{\pmb{k}}(\omega-\omega_{\pmb{k}})$  and $(k\pm2\varkappa)^2\simeq2\omega_{\pmb{k}\pm2\pmb{\varkappa}}(\omega-\omega_{\pmb{k}\pm2\pmb{\varkappa}}\pm2\varkappa_0)$.
Observe that this last linearization applies whenever  the  conditions $k\varkappa\simeq0$ and $\omega>2\varkappa_0$ are fulfilled. In correspondence,
one can deal with a simplified version of the eigenproblems given in
Eqs.~(\ref{eigenproblem})-(\ref{statesflavornomass}):
\begin{equation}\label{dispr1va}
\underbrace{\left[
\begin{array}{ccc}
\omega_{\pmb{k}}-\omega &-\frac{\pi_0(k)}{\omega_{\pmb{k}}}\\
-\frac{\pi_0(k)}{\omega_{\pmb{k}+2\pmb{\varkappa}}} & \omega_{\pmb{k}+2\pmb{\varkappa}}-2\varkappa_0-\omega
\end{array}\right]}_{\pmb{\mathcal{G}}^{(1)}}\underbrace{\left[\begin{array}{c} f_+(k)\\ f_-(k+2\varkappa)
\end{array}\right]}_{\pmb{\mathpzc{z}}^{(1)}}=0,\qquad \underbrace{\left[
\begin{array}{ccc}
\omega_{\pmb{k}-2\pmb{\varkappa}}+2\varkappa_0-\omega &-\frac{\pi_0(k)}{\omega_{\pmb{k}-2\pmb{\varkappa}}}\\
-\frac{\pi_0(k)}{\omega_{\pmb{k}}} & \omega_{\pmb{k}}-\omega
\end{array}\right]}_{\pmb{\mathcal{G}}^{(2)}}\underbrace{\left[\begin{array}{c} f_+(k-2\varkappa)\\ f_-(k)
\end{array}\right]}_{\pmb{\mathpzc{z}}^{(2)}}=0
\end{equation} The respective eigenvalues  are well approached by $\omega_1^{(1,2)}\simeq\omega_{\pmb{k}}$ and $\omega_2^{(1,2)}\simeq\omega_{\pmb{k}\pm2\pmb{\varkappa}}\mp2\varkappa_0$,
whereas the corresponding  eigenvectors are
\begin{equation}
\pmb{\mathpzc{z}}_1^{(1,2)}=\frac{1}{\left[1+\tan^2\left(\varphi_1^{(1,2)}\right)\right]^{\nicefrac{1}{2}}}\left[\begin{array}{c} 1 \\-\tan\left(\varphi_1^{(1,2)}\right)
\end{array}\right]\qquad\mathrm{and}\qquad\pmb{\mathpzc{z}}_2^{(1,2)}=\frac{1}{\left[1+\tan^2\left(\varphi_2^{(1,2)}\right)\right]^{\nicefrac{1}{2}}}\left[\begin{array}{c} \tan\left(\varphi_2^{(1,2)}\right)\\1
\end{array}\right].\label{normalizeeigenstates}
\end{equation}Note that the upper indices  are  used to distinguish the quantities associated with each eigenproblem. We find convenient to emphasize that these
eigenstates have been calculated with accuracy of terms  $\sim \mathpzc{o}(\alpha_\epsilon^2)$ and  turn out to be  parameterized by the small
angles $\varphi_{i}^{(1,2)}\ll1$ with ($i=1,2$) and
\begin{eqnarray}
\varphi_1^{(1,2)}=-\left.\frac{f_\mp(k\pm2\varkappa)}{f_\pm(k)}\right\vert_{\omega=\omega_1^{(1,2)}}=\frac{\pi_0(k)}{\omega_{\pmb{k}\pm2\pmb{\varkappa}}(\omega_{\pmb{k}}-\omega_{\pmb{k}\pm2\pmb{\varkappa}}\pm2\varkappa_0)},\quad
\varphi_2^{(1,2)}=\left.\frac{f_\pm(k)}{f_\mp(k\pm2\varkappa)}\right\vert_{\omega=\omega_2^{(1,2)}}=\frac{\pi_0(k)}{\omega_{\pmb{k}}(\omega_{\pmb{k}}-\omega_{\pmb{k}\pm2\pmb{\varkappa}}\pm2\varkappa_0)}.
\end{eqnarray}

The  previous linearizations in the dispersion equations  are equivalent to  reduce  the differential order in the  equation of motion [Eq.~(\ref{sdpmBF})]. In correspondence
one  can approach  the outgoing  state   $\pmb{\mathpzc{z}}^{(1,2)}$ as a superposition of the two mass eigenstates  that  characterize the process
\begin{equation}\label{phasespacesolution}
\pmb{\mathpzc{z}}^{(1,2)}(\omega)\simeq\sum_{\lambda=1,2}\mathcal{B}_\lambda^{(1,2)} \pmb{\mathpzc{z}}_{\lambda}^{(1,2)}\delta \left(\omega-\omega_\lambda^{(1,2)}\right).
\end{equation}Here  $\mathcal{B}_\lambda^{(1,2)}$ denote  some constants to be determined by the initial conditions.  For the sake  of a better understanding, we Fourier transform
Eq.~(\ref{phasespacesolution}) only in time. Next,  we  consider  the experimental setup in  which the
incoming  probe beam is  a linear combination of circularly polarized  waves with opposite helicities  [Eq.~(\ref{chiralityexpansio})]. Besides, we  suppose  that  at
$t=0$  only the  incoming  beam has a nonvanishing amplitude with  $f_{\pm}(\pmb{k},0)=[4\pi/(2\omega_{\pmb{k}})]^{\nicefrac{1}{2}}.$ Following this procedure, one  obtains
a system of algebraic equations for $\mathcal{B}_\lambda^{(1,2)}$.   Its solution allows to  approach  the  components of the outgoing electromagnetic wave by
\begin{eqnarray}\label{afield}
f_\pm(\pmb{k},t)={\sqrt{\frac{4\pi}{2\omega_{\pmb{k}}}}\mathpzc{A}_\pm(\pmb{k},t)}e^{-i\omega_{\pmb{k}}t}\quad\mathrm{and}\quad
f_\mp(\pmb{k}\pm2\pmb{\varkappa},t)=\sqrt{\frac{4\pi}{2\omega_{\pmb{k}\pm2\pmb{\varkappa}}}}\mathpzc{A}_\mp(\pmb{k}\pm2\pmb{\varkappa},t)e^{-i\omega_{\pmb{k}\pm2\pmb{\varkappa}}t}.
\end{eqnarray} The amplitudes contained in the expressions above read
\begin{eqnarray}\label{resonantamplitudee}
\begin{array}{c}\displaystyle
\mathpzc{A}_\pm(\pmb{k},t)\approx \exp\left\{-i\varphi_1^{(1,2)}\varphi_2^{(1,2)}\sin\left[\left(\omega_{\pmb{k}}-\omega_{\pmb{k}\pm 2\pmb{\varkappa}}\pm 2\varkappa_0\right)t\right]-2\varphi_1^{(1,2)}\varphi_2^{(1,2)}\sin^2\left[\frac{1}{2}\left(\omega_{\pmb{k}}-\omega_{\pmb{k}\pm2\pmb{\varkappa}}\pm2\varkappa_0\right)t\right]\right\},\\ \\
\displaystyle \mathpzc{A}_\mp(\pmb{k}\pm2\pmb{\varkappa},t)\approx -\varphi_1^{(1,2)}\sqrt{\frac{\omega_{\pmb{k}\pm2\pmb{\varkappa}}}{\omega_{\pmb{k}}}}\left\{2\sin^2\left[\frac{1}{2}\left(\omega_{\pmb{k}}-\omega_{\pmb{k}\pm2\pmb{\varkappa}}\pm2\varkappa_0\right)t\right]-i\sin\left[\left(\omega_{\pmb{k}}-\omega_{\pmb{k}\pm2\pmb{\varkappa}}\pm2\varkappa_0\right)t\right]\right\}.
\end{array}
\end{eqnarray} Clearly, the  square of $\mathpzc{A}_\mp(\pmb{k}\pm2\pmb{\varkappa},t)$ provides the photo-production probability of a Raman-like photon.
The resulting  expression  is intrinsically associated with the  exponentials responsible for the damping  of the corresponding electromagnetic wave due to
the   mixing  of  photons  with  different helicities [second term in the exponent of $\mathpzc{A}_{\pm}(\pmb{k},t)$].  We combine  the respective outcomes
to    express   the total photo-production  probability of Raman-like
waves as
\begin{equation}
\mathpzc{P}_{\gamma\to\gamma^\prime}(t)=\frac{4 \vert\pi_0(k)\vert^2}{\omega_{\pmb{k}+2\pmb{\varkappa}}\omega_{\pmb{k}}}\frac{\sin^2\left[\frac{1}{2}\left(\omega_{\pmb{k}}-\omega_{\pmb{k}+2\pmb{\varkappa}}+2\varkappa_0\right)t\right]}{\left(\omega_{\pmb{k}}-\omega_{\pmb{k}+2\pmb{\varkappa}}+2\varkappa_0\right)^2}+\frac{4\vert\pi_0(k)\vert^2}{\omega_{\pmb{k}-2\pmb{\varkappa}}\omega_{\pmb{k}}}\frac{\sin^2\left[\frac{1}{2}\left(\omega_{\pmb{k}}-\omega_{\pmb{k}-2\pmb{\varkappa}}-2\varkappa_0\right)t\right]}{\left(\omega_{\pmb{k}}-\omega_{\pmb{k}-2\pmb{\varkappa}}-2\varkappa_0\right)^2}\label{progene}.
\end{equation} It is worth mentioning at this point that $\lim_{t\to\infty}\mathpzc{P}_{\gamma\to\gamma^\prime}(t)/t=\mathpzc{R}_{\gamma\to\gamma^\prime}$
reproduces  Eq.~(\ref{ratefermionphotongeneration}). This statement can be verified by considering the relation $\pi\delta(x)=\lim_{\tau\to\infty}\sin^2(x\tau)/(x^2\tau)$.

We wish to particularize Eq.~(\ref{progene}) to the case in which  both lasers propagate quasi-parallelly, i.e.,   when  $k\varkappa\approx\omega\varkappa_0\theta^2/2\ll1$ with $\theta$
denoting the collision angle [$\theta\ll1$]. In this framework, the  conversion probability, resulting from the substitution of Eq.~(\ref{imprt}) into Eq.~(\ref{progene}), is given by
\begin{eqnarray}\label{strongfieldramman}
\begin{array}{c}\displaystyle
\mathpzc{P}_{\gamma\to\gamma^\prime}=\mathpzc{P}_{\omega\to\omega+2\varkappa_0}+\mathpzc{P}_{\omega\to\omega-2\varkappa_0},\\ \\
\displaystyle
\mathpzc{P}_{\omega\to\omega\pm2\varkappa_0}\approx\epsilon^{\nicefrac{16}{3}}\frac{\alpha^2  \xi^{\nicefrac{4}{3}}\Gamma^4\left(\frac{2}{3}\right)}{42^2 \pi\ \lambda^{\nicefrac{2}{3}} \ \Gamma^2\left(\frac{7}{6}\right)}\left(\frac{2}{3}\right)^{\nicefrac{4}{3}}\left\vert 1\pm2\frac{\varkappa_0}{\omega}\right\vert\sin^2\left[\frac{2m^2\lambda}{\omega\pm2\varkappa_0}t\right].
\end{array}
\end{eqnarray}We find  opportune to emphasize that  Eq.~(\ref{strongfieldramman}) applies for both $\omega>2\varkappa_0$ or $2\varkappa_0>\omega$.
Moreover,  it is  valid  whenever the condition $m_\epsilon\ll m\left[\epsilon\lambda\right]^{\nicefrac{1}{3}}$ is fulfilled.  Here   the parameter
$\lambda$ must be understood as  $\lambda\approx\omega\varkappa_0\theta^2/(4m^2)$.

\subsection{Raman spectroscopy   as a probe of MCPs  \label{rammanwavespoltensor}}

Now that Eq.~(\ref{progene}) has been  established we briefly provide some details about the  experimental configuration.
The nature of the   waves produced in the inelastic process shares certain similarities with Raman-dispersion in solid-state physics. Therefore, in the
search  of constraints  on the MCPs it would be convenient to exploit the well known techniques of the Raman's spectroscopy.  So, we  suppose  that  after
the interaction with the strong field of the high-intensity laser,  the outgoing probe electromagnetic wave  is  picked up with a lens and sent to a
monochromator.  The latter device  allows us to filter out the part of the probe beam which is elastically scattered and, in correspondence,  only those
photons with  frequency $\omega+2\varkappa_0$ or $\omega-2\varkappa_0$    are analyzed in a detector.

Let us  consider  the search of Raman's photons with $\omega+2\varkappa_0$. We  suppose the situation in which the  collision occurs with an angle  $\theta\simeq10^\circ$.
Our calculations will be initially  particularized
with the envisaged  parameters of  the  POLARIS system \cite{POLARIS} [$\xi\sim 10^2$, $\varkappa_0=1.2\ \rm eV$ and  $\tau_{\mathrm{p}}=120\ \rm fs$]. For the  probe beam, we employ the multi-TW
class laser JETI\footnote{The feasibility of  this experimental setup has   been    theoretically  exploited  in the search of Axion-like particles \cite{Dobrich:2010hi,Dobrich:2010ie}.},
which--after a second upgrade--could deliver  up to $3 \ \rm J$ per shot in a  pulse  length  $\tau_{\mathrm{J}}\simeq 30\ \rm fs$  at  frequency $\omega=1.55 \ \rm eV$.
Accordingly,  the number of probe photons  emitted per shot might reach $\mathpzc{N}_0\simeq 1.21\times 10^{19}$. In our case, the  excluded regions on the
$(m_\epsilon,\epsilon)$-plane are then  settled  by requiring  a single-Raman's photon detection for $\dot{\mathpzc{N}}$. This fact allows us to claim
$\dot{\mathpzc{N}}/\dot{\mathpzc{N}}_0>\mathpzc{P}_{\gamma\to\gamma^\prime}$. In such a case, $\dot{\mathpzc{N}}/\dot{\mathpzc{N}}_{0}\approx 8.3\times10^{-20}$
could be established  and  the upper bound  $\epsilon<6.5\times 10^{-5}$ is found  for $m_\epsilon\ll 3.4\ \rm eV$. This constraint  is comparable with  those
obtained from a polarimetric search when both lasers counterpropagate [see Sec. \ref{perspective}]. Let us consider the case in which  the total measurement time
 is one year. Since POLARIS has a repetition rate $f_{\mathrm{rep}}\simeq0.1\ \mathrm{Hz}$--leading in practice to $\mathpzc{o}(100)$ shots per day-- one can establish  the
 upper bound $\epsilon<9.1\times 10^{-6}$ for masses much below the eV-regime.

The situation could be  more stringent when  the envisaged experimental designations associated  with the  ELI and  XCELS projects are considered. In these ultra-high
intensity laser systems, a power $P\approx 1\ \rm EW$, with  $\xi\approx 6.7\times 10^3$ and central  frequency $\varkappa_0\simeq1.55\ \rm eV$ is planned. The combination
of the latter  with the temporal extension $\tau\simeq 15 \ \rm fs$  gives us  $\varkappa_0\tau\approx 35$. Obviously,  the monochromaticity condition is not as well satisfied
as  in the POLARIS case. Nonetheless, some interesting estimations can be done. For instance, by  keeping   the collision angle $\theta\simeq10^\circ$ and under the  assumption
of a single-Raman's  photon detection, it is found that an  upper bound--like the  best laboratory  based one $\epsilon< 5\times 10^{-7}$ \cite{Ehret:2010mh}--would require
an optical probe source delivering $\mathpzc{N}_0\sim 4 \times 10^{29}$ photons per shot. Although  the latter requirement is far from the capability of the existing  facilities, the
fast development  of laser technology  offers prospects  that it can be reached--even overpassed--in a near future.

\section{Summary, discussion and outlook}

Vacuum polarization effects induced by the interaction of  MCPs and  a high-intensity laser wave  provide  alternative scenarios
for  probing some low-energy effective SM extensions in which such  hypothetical particles are included. In this work we have focused ourselves to
the particular  situation  where the strong laser field  is  circularly polarized. We  have found that in some asymptotic limits,  the birefringence and dichroism of
the vacuum  are less pronounced than in the case in which the polarization is driven by a constant field. In particular, this holds in a region far from the threshold
of pair production. Certainly, this situation is not favorable  in the search of  MCPs when the polarimetric techniques, with an ultra-high-intensity  laser,  are  thought
as the main experimental tools to be implemented. Nonetheless, evidences resulting
from   an effective Lagrangian  treatment  reveal  a  strong birefringent and dichroic character of  the vacuum as the   strong field  of
the wave is, for instance, linearly polarized.  Therefore, much more  severe  constraints   could   arise.  The problem, however, becomes more
cumbersome  because the form factors of  $\Pi_{\mu\nu}$  are strongly dependent on Bessel functions \cite{baier,VillalbaChavez:2012bb,baierbook}.
Yet, the possibility of exploiting the quasi-static limit in the strong field regime  has put forward interesting estimations
\cite{Gies:2008wv}.

In a vicinity of the region in which the photo-production of a pair occurs, the birefringent and dichroic  properties of the vacuum
are quite pronounced. Both phenomena are closely connected with the chiral activity of the ``medium'' and could be observed even at
intensities available today.  Observation of these elusive effects would provide evidences on the nonlinear feature of the QED vacuum.
In addition, they would complement our understanding of the multi-photon pair  production,  already detected using  nonlinear Compton
scattering in the SLAC E144 experiment \cite{Burke:1997ew}.   Moreover, at  such  external field strengths,  the search  of  MCPs
by using high-precision polarimetric  experiments is suitable and could provide new constraints  on  $\epsilon$  in regions of masses
where the searches  based on dipole  magnets are less stringent.  We have  shown  that the latter  statement applies for Dirac but also
for Klein-Gordon representations of such hypothetical charge carriers.
Finally, in the last part of this work,  the generation of small-amplitude electromagnetic waves  resulting  from the inelastic part  of the photon-photon
scattering was investigated. We have noted  that  Raman's spectroscopy in a vacuum polarized by a high-intensity circular polarized  laser wave
could provide a sensitive  probe of MCPs as well.  Parameters of modern laser systems   were  used for establishing  upper bounds on
the parameters of MCPs.

\vspace{0.005 in}
\begin{flushleft}
\textbf{Acknowledgments}
\end{flushleft}
\vspace{0.005 in}
S. Villalba-Chavez thanks Babette D\"{o}brich for helpful discussions.  He also gratefully acknowledges the support by the Alexander von Humboldt Foundation.


\begin{thebibliography}{28}

\expandafter\ifx\csname
natexlab\endcsname\relax\def\natexlab#1{#1}\fi
\expandafter\ifx\csname bibnamefont\endcsname\relax
  \def\bibnamefont#1{#1}\fi
\expandafter\ifx\csname bibfnamefont\endcsname\relax
  \def\bibfnamefont#1{#1}\fi
\expandafter\ifx\csname citenamefont\endcsname\relax
  \def\citenamefont#1{#1}\fi
\expandafter\ifx\csname url\endcsname\relax
  \def\url#1{\texttt{#1}}\fi
\expandafter\ifx\csname urlprefix\endcsname\relax\def\urlprefix{URL
}\fi \providecommand{\bibinfo}[2]{#2}
\providecommand{\eprint}[2][]{\url{#2}}

\bibitem{Weinberg:2000} \bibinfo{author}{\bibfnamefont{S.}~%
\bibnamefont{Weinberg}}. \emph{%
\bibinfo{journal}{``The Quantum theory of fields.''}} \pmb{\bibinfo{volume}{III}}, \bibinfo{editor}{Cambridge, UK: Univ. Pr.}, (\bibinfo{year}{2000}%
), 441 p.

\bibitem{polchinski} \bibinfo{author}{\bibfnamefont{J.}~%
\bibnamefont{Polchinski}}. \emph{
\bibinfo{journal}{``String theory.''}} Vol. \pmb{\bibinfo{volume}{I}} and \pmb{\bibinfo{volume}{II}} , \bibinfo{editor}{Cambridge, UK: Univ. Pr.}, (\bibinfo{year}{2001,\ 2005}).

\bibitem{Jaeckel:2010ni}
\bibinfo{author}{\bibfnamefont{J.}~\bibnamefont{Jaeckel}}  \bibnamefont{and}
\bibinfo{author}{\bibfnamefont{A.}~\bibnamefont{Ringwald}}.
\bibinfo{journal}{ Ann.\ Rev.\ Nucl.\ Part.\ Sci.} \pmb{\bibinfo{volume}{60}},
\bibinfo{pages}{405} (\bibinfo{year}{2010}); [arXiv:1002.0329 [hep-ph]].

\bibitem{Redondo:2010dp}
\bibinfo{author}{\bibfnamefont{J.}~\bibnamefont{Redondo}}  \bibnamefont{and}
\bibinfo{author}{\bibfnamefont{A.}~\bibnamefont{Ringwald}}.
\bibinfo{journal}{Contemp.\ Phys.\ } \pmb{\bibinfo{volume}{52}},
\bibinfo{pages}{211} (\bibinfo{year}{2011}); [arXiv:1011.3741 [hep-ph]].

\bibitem{Gies:2007ua}
\bibinfo{author}{\bibfnamefont{H.}~\bibnamefont{Gies}}.
\bibinfo{journal}{J.\ Phys.\ A} \pmb{\bibinfo{volume}{41}},
\bibinfo{pages}{164039} (\bibinfo{year}{2008});  [arXiv:0711.1337 [hep-ph]].

\bibitem{Gies:2008wv}
\bibinfo{author}{\bibfnamefont{H.}~\bibnamefont{Gies}}.
\bibinfo{journal}{Eur. Phys. J.  D} \pmb{\bibinfo{volume}{55}},
\bibinfo{pages}{311} (\bibinfo{year}{2009}); [arXiv:0812.0668 [hep-ph]].

\bibitem{Witten:1984dg}
\bibinfo{author}{\bibfnamefont{E.}~\bibnamefont{Witten}}.
\bibinfo{journal}{ Phys.\ Lett.\ B } \pmb{\bibinfo{volume}{149}},
\bibinfo{pages}{351} (\bibinfo{year}{1984}).

 \bibitem{Lebedev:2009ag}
\bibinfo{author}{\bibfnamefont{O.} \bibnamefont{Lebedev}}  \bibnamefont{and}
\bibinfo{author}{\bibfnamefont{S.~Ramos}~\bibnamefont{Sanchez}}.
\bibinfo{journal}{ Phys.\ Lett.\ B} \pmb{\bibinfo{volume}{684}},
\bibinfo{pages}{48} (\bibinfo{year}{2010}); [arXiv:0912.0477 [hep-ph]].

\bibitem{Okun:1982xi}
\bibinfo{author}{\bibfnamefont{L.~B.}~\bibnamefont{Okun}}.
\bibinfo{journal}{Sov. Phys. JETP} \pmb{\bibinfo{volume}{56}},
\bibinfo{pages}{502} (\bibinfo{year}{1982});  [Zh.\ Eksp.\ Teor.\ Fiz.\  {\bf 83} (1982) 892].

\bibitem{Ahlers:2007rd}
\bibinfo{author}{\bibfnamefont{M.}~\bibnamefont{Ahlers}},
\bibinfo{author}{\bibfnamefont{H.}~\bibnamefont{Gies}},
\bibinfo{author}{\bibfnamefont{J.}~\bibnamefont{Jaeckel}},
\bibinfo{author}{\bibfnamefont{J.}~\bibnamefont{Redondo}}
\bibnamefont{and}
\bibinfo{author}{\bibfnamefont{A.}~\bibnamefont{Ringwald}}.
\bibinfo{journal}{Phys. Rev. D.} \pmb{\bibinfo{volume}{76}},
\bibinfo{pages}{115005} (\bibinfo{year}{2007});  [arXiv:0706.2836 [hep-ph]].

\bibitem{Ahlers:2007qf}
\bibinfo{author}{\bibfnamefont{M.}~\bibnamefont{Ahlers}},
\bibinfo{author}{\bibfnamefont{H.}~\bibnamefont{Gies}},
\bibinfo{author}{\bibfnamefont{J.}~\bibnamefont{Jaeckel}},
\bibinfo{author}{\bibfnamefont{J.}~\bibnamefont{Redondo}},
\bibnamefont{and}
\bibinfo{author}{\bibfnamefont{A.}~\bibnamefont{Ringwald}}.
\bibinfo{journal}{Phys. Rev. D.} \pmb{\bibinfo{volume}{77}},
\bibinfo{pages}{ 095001} (\bibinfo{year}{2008});  [arXiv:0711.4991 [hep-ph]].

\bibitem{Goodsell:2009xc}
\bibinfo{author}{\bibfnamefont{M.}~\bibnamefont{Goodsell}},
\bibinfo{author}{\bibfnamefont{J.}~\bibnamefont{Jaeckel}},
\bibinfo{author}{\bibfnamefont{J.}~\bibnamefont{Redondo}}
\bibnamefont{and}
\bibinfo{author}{\bibfnamefont{A.}~\bibnamefont{Ringwald}}.
\bibinfo{journal}{JHEP} \pmb{\bibinfo{volume}{0911}},
\bibinfo{pages}{027} (\bibinfo{year}{2009}); [arXiv:0909.0515 [hep-ph]].

\bibitem{Dudas:2012pb}
\bibinfo{author}{\bibfnamefont{E.}~\bibnamefont{Dudas}},
\bibinfo{author}{\bibfnamefont{Y.}~\bibnamefont{Mambrini}},
\bibinfo{author}{\bibfnamefont{S.}~\bibnamefont{Pokorski}}
\bibnamefont{and}
\bibinfo{author}{\bibfnamefont{A.}~\bibnamefont{Romagnoni}}.
\bibinfo{journal}{JHEP} \pmb{\bibinfo{volume}{1210}},
\bibinfo{pages}{123} (\bibinfo{year}{2012}); [arXiv:1205.1520 [hep-ph]].

\bibitem{Holdom:1985ag}
\bibinfo{author}{\bibfnamefont{B.}~\bibnamefont{Holdom}}.
\bibinfo{journal}{Phys. Lett. B } \pmb{\bibinfo{volume}{166}},
\bibinfo{pages}{196} (\bibinfo{year}{1986}).

\bibitem{Masso:2006gc}
\bibinfo{author}{\bibfnamefont{E.}~\bibnamefont{Masso}}  \bibnamefont{and}
\bibinfo{author}{\bibfnamefont{J.}~\bibnamefont{Redondo}}.
\bibinfo{journal}{Phys. Rev. Lett.} \pmb{\bibinfo{volume}{97}},
\bibinfo{pages}{151802} (\bibinfo{year}{2006});  [arXiv:hep-ph/0606163].

\bibitem{Gies:2006ca}
\bibinfo{author}{\bibfnamefont{H.}~\bibnamefont{Gies}},
\bibinfo{author}{\bibfnamefont{J.}~\bibnamefont{Jaeckel}}
\bibnamefont{and}
\bibinfo{author}{\bibfnamefont{A.}~\bibnamefont{Ringwald}}.
\bibinfo{journal}{Phys. Rev. Lett.} \pmb{\bibinfo{volume}{97}},
\bibinfo{pages}{140402} (\bibinfo{year}{2006}); [arXiv:hep-ph/0607118].

\bibitem{Jaeckel:2009dh}
\bibinfo{author}{\bibfnamefont{J.}~\bibnamefont{Jaeckel}}.
\bibinfo{journal}{Phys. Rev. Lett.} \pmb{\bibinfo{volume}{103}},
\bibinfo{pages}{080402} (\bibinfo{year}{2009}); [arXiv:0904.1547 [hep-ph]].

\bibitem{Bruemmer:2009ky}
\bibinfo{author}{\bibfnamefont{F.}~\bibnamefont{Brummer}},
\bibinfo{author}{\bibfnamefont{J.}~\bibnamefont{Jaeckel}}
\bibnamefont{and}
\bibinfo{author}{\bibfnamefont{V.~V.}~\bibnamefont{Khoze}}.
\bibinfo{journal}{JHEP} \pmb{\bibinfo{volume}{0906}},
\bibinfo{pages}{037} (\bibinfo{year}{2009}); [arXiv:0905.0633 [hep-ph]].

\bibitem{Cameron:1993mr}
\bibinfo{author}{\bibfnamefont{R.}~\bibnamefont{Cameron}  {\it et al.}}
\bibinfo{journal}{Phys.  Rev.  D } \pmb{\bibinfo{volume}{47}},
\bibinfo{pages}{3707 } (\bibinfo{year}{1993}).

\bibitem{Zavattini:2007ee}
\bibinfo{author}{\bibfnamefont{E.}~\bibnamefont{Zavattini}  {\it et al.}  [PVLAS Collaboration]}.
\bibinfo{journal}{Phys.  Rev.  D } \pmb{\bibinfo{volume}{77}},
\bibinfo{pages}{032006} (\bibinfo{year}{2008}).

\bibitem{BMVreport}
\bibinfo{author}{\bibfnamefont{R.}~\bibfnamefont{Battesti}  {\it et al.}}
\bibinfo{journal}{Eur. Phys. J.  D} \pmb{\bibinfo{volume}{46}},
\bibinfo{pages}{323} (\bibinfo{year}{2008}).

\bibitem{Chen:2006cd}
\bibinfo{author}{\bibfnamefont{S.}~\bibfnamefont{J.}~\bibnamefont{Chen} {\it et al.}}
\bibinfo{journal}{Mod. Phys. Lett.  A} \pmb{\bibinfo{volume}{22}},
\bibinfo{pages}{2815} (\bibinfo{year}{2007}).

\bibitem{Dittrich}
\bibinfo{author}{\bibfnamefont{W.~} \bibnamefont{Dittrich}} \bibnamefont{and}
\bibinfo{author}{\bibfnamefont{H.~} \bibnamefont{Gies}}.
\bibinfo{editor}{Springer, Heidelberg}, (\bibinfo{year}{2000}).

\bibitem{adler} \bibinfo{author}{\bibfnamefont{S.}~\bibnamefont{L.}~%
\bibnamefont{Adler}},
\bibinfo{author}{\bibfnamefont{J.~N.}
\bibnamefont{Bahcall}},
\bibinfo{author}{\bibfnamefont{C.~G.}
\bibnamefont{Callan}} and
\bibinfo{author}{\bibfnamefont{M.~N.}
\bibnamefont{Rosenbluth}}.
\bibinfo{journal}{Phys. Rev. Lett.} \pmb{\bibinfo{volume}{25}}, %
\bibinfo{pages}{1061} (\bibinfo{year}{1970})

\bibitem{shabad4}
\bibinfo{author}{\bibfnamefont{A.~E.} \bibnamefont{Shabad}}.
\bibinfo{journal}{Sov. Phys. JETP} \pmb{\bibinfo{volume}{98}},
\bibinfo{pages}{186} (\bibinfo{year}{2004}).

\bibitem{VillalbaChavez:2012ea}
\bibinfo{author}{\bibfnamefont{S.}~\bibnamefont{Villalba-Chavez}}
\bibnamefont{and}
\bibinfo{author}{\bibfnamefont{A.~E.~} \bibnamefont{Shabad}}.
\bibinfo{journal}{Phys.\  Rev.\  D} \pmb{\bibinfo{volume}{86}},
\bibinfo{pages}{105040} (\bibinfo{year}{2012});  arXiv:1206.4491 [hep-th].

\bibitem{Chavez:2009ia}
\bibinfo{author}{\bibfnamefont{S.}~\bibnamefont{Villalba}~\bibnamefont{Ch\'avez}}.
\bibinfo{journal}{Phys. Rev. D} \pmb{\bibinfo{volume}{81}},
\bibinfo{pages}{105019}, (\bibinfo{year}{2010}); arXiv:0910.5149 [hep-th].

\bibitem{Hattori}
\bibinfo{author}{\bibfnamefont{K.}~\bibnamefont{Hattori}}
\bibnamefont{and}
\bibinfo{author}{\bibfnamefont{K.}~\bibnamefont{Itakura}}.
\bibinfo{journal}{Ann.\  Phys.} \pmb{\bibinfo{volume}{330}},
\bibinfo{pages}{23} (\bibinfo{year}{2013}).

\bibitem{Chou:2007zzc}
\bibinfo{author}{\bibfnamefont{A.}~\bibfnamefont{S.}~\bibfnamefont{Chou}  {\it et al.}  [GammeV (T-969) Collaboration].}
\bibinfo{journal}{Phys. Rev. Lett.} \pmb{\bibinfo{volume}{100}},
\bibinfo{pages}{080402} (\bibinfo{year}{2008});  [arXiv:0710.3783 [hep-ex]].

\bibitem{Steffen:2009sc}
\bibinfo{author}{\bibfnamefont{J.}~\bibnamefont{H.}~\bibnamefont{Steffen}}
\bibnamefont{and}
\bibinfo{author}{\bibfnamefont{A.}~\bibnamefont{Upadhye}}.
\bibinfo{journal}{Mod. Phys. Lett. A} \pmb{\bibinfo{volume}{24}},
\bibinfo{pages}{2053} (\bibinfo{year}{2009}); [arXiv:0908.1529 [hep-ex]].

\bibitem{Afanasev:2008jt}
\bibinfo{author}{\bibfnamefont{A.}~\bibfnamefont{Afanasev}  {\it et al.}}
\bibinfo{journal}{Phys. Rev.  Lett.} \pmb{\bibinfo{volume}{101}},
\bibinfo{pages}{120401} (\bibinfo{year}{2008});  [arXiv:0806.2631 [hep-ex]].

\bibitem{Pugnat:2007nu}
\bibinfo{author}{\bibfnamefont{P.}~\bibfnamefont{Pugnat}  {\it et al.} [OSQAR Collaboration].}
\bibinfo{journal}{Phys. Rev. D} \pmb{\bibinfo{volume}{78}},
\bibinfo{pages}{092003} (\bibinfo{year}{2008}); [arXiv:0712.3362 [hep-ex]].

\bibitem{Robilliard:2007bq}
\bibinfo{author}{\bibfnamefont{C.}~\bibnamefont{Robilliard}} {\it et al.}.
\bibinfo{journal}{Phys.  Rev.  Lett.} \pmb{\bibinfo{volume}{99}},
\bibinfo{pages}{190403} (\bibinfo{year}{2007}); [arXiv:0707.1296 [hep-ex]].

\bibitem{Fouche:2008jk}
\bibinfo{author}{\bibfnamefont{M.}~\bibfnamefont{~Fouche}  {\it et al.}}
\bibinfo{journal}{Phys.  Rev.  D.} \pmb{\bibinfo{volume}{78}},
\bibinfo{pages}{032013} (\bibinfo{year}{2008}); [arXiv:0808.2800 [hep-ex]].

\bibitem{Ehret:2010mh}
\bibinfo{author}{\bibfnamefont{K.}~\bibfnamefont{Ehret}  {\it et al.}  [ALPS collaboration]}
\bibinfo{journal}{Phys. Lett.   B } \pmb{\bibinfo{volume}{689}},
\bibinfo{pages}{149} (\bibinfo{year}{2010}); [arXiv:1004.1313 [hep-ex]].

\bibitem{Ehret:2009sq}
\bibinfo{author}{\bibfnamefont{K.}~\bibfnamefont{Ehret}  {\it et al.}  [ALPS collaboration].}
\bibinfo{journal}{Nucl. Instrum. Meth.  A} \pmb{\bibinfo{volume}{612}},
\bibinfo{pages}{83} (\bibinfo{year}{2009}); [arXiv:0905.4159 [physics.ins-det]].

\bibitem{VanBibber:1987rq}
\bibinfo{author}{\bibfnamefont{K.}~\bibfnamefont{Van}~\bibnamefont{Bibber}},
\bibinfo{author}{\bibfnamefont{N.~R.}~\bibnamefont{Dagdeviren}},
\bibinfo{author}{\bibfnamefont{S.~E.}~\bibnamefont{Koonin}},
\bibinfo{author}{\bibfnamefont{A.}~\bibnamefont{Kerman}}  \bibnamefont{and}
\bibinfo{author}{\bibfnamefont{H.~N.}~\bibnamefont{Nelson}}.
\bibinfo{journal}{Phys. Rev. Lett.} \pmb{\bibinfo{volume}{59}},
\bibinfo{pages}{759} (\bibinfo{year}{1987}).

\bibitem{Adler:2008gk}
\bibinfo{author}{\bibfnamefont{S.~L.}~\bibnamefont{Adler}},
\bibinfo{author}{\bibfnamefont{J.}~\bibnamefont{Gamboa}},
\bibinfo{author}{\bibfnamefont{F.}~\bibnamefont{Mendez}}  \bibnamefont{and}
\bibinfo{author}{\bibfnamefont{J.}~\bibnamefont{Lopez-Sarrion}}.
\bibinfo{journal}{Ann. Phys.} \pmb{\bibinfo{volume}{323}},
\bibinfo{pages}{2851} (\bibinfo{year}{2008}); [arXiv:0801.4739 [hep-ph]].

\bibitem{Arias:2010bh}
\bibinfo{author}{\bibfnamefont{P.}~\bibnamefont{Arias}},
\bibinfo{author}{\bibfnamefont{J.}~\bibnamefont{Jaeckel}}  \bibnamefont{and}
\bibinfo{author}{\bibfnamefont{A.}~\bibnamefont{Ringwald}}.
\bibinfo{journal}{Phys. Rev.  D } \pmb{\bibinfo{volume}{82}},
\bibinfo{pages}{115018} (\bibinfo{year}{2010}); [arXiv:1009.4875 [hep-ph]].

\bibitem{Dobrich:2012sw}
\bibinfo{author}{\bibfnamefont{B.}~\bibnamefont{D\"obrich}},
\bibinfo{author}{\bibfnamefont{H.}~\bibnamefont{Gies}},
\bibinfo{author}{\bibfnamefont{N.}~\bibnamefont{Neitz}}
\bibnamefont{and}
\bibinfo{author}{\bibfnamefont{F.}~\bibnamefont{Karbstein}}.
\bibinfo{journal}{ Phys.\ Rev.\ Lett.\ } \pmb{\bibinfo{volume}{109}},
\bibinfo{pages}{131802} (\bibinfo{year}{2012}); [arXiv:1203.2533 [hep-ph]].

\bibitem{Dobrich:2012jd}
\bibinfo{author}{\bibfnamefont{B.}~\bibnamefont{D\"obrich}},
\bibinfo{author}{\bibfnamefont{H.}~\bibnamefont{Gies}},
\bibinfo{author}{\bibfnamefont{N.}~\bibnamefont{Neitz}}
\bibnamefont{and}
\bibinfo{author}{\bibfnamefont{F.}~\bibnamefont{Karbstein}}.
\bibinfo{journal}{ Phys.\ Rev.\ D  } \pmb{\bibinfo{volume}{87}},
\bibinfo{pages}{025022} (\bibinfo{year}{2013});  [arXiv:1203.4986 [hep-ph]].

\bibitem{ELI} See: http://www.extreme-light-infrastructure.eu

\bibitem{xcels} See: http://www.xcels.iapras.ru/

\bibitem{baier}
\bibinfo{author}{\bibfnamefont{V.~N.~} \bibnamefont{Ba\u{\i}er}},
\bibinfo{author}{\bibfnamefont{A.~I.~} \bibnamefont{Mil'shte\u{\i}n}}
\bibnamefont{and}
\bibinfo{author}{\bibfnamefont{V.~M.~} \bibnamefont{Strakhovenko}}.
\bibinfo{journal}{Zh. Eksp. Teo. Fiz.} \pmb{\bibinfo{volume}{69}},
\bibinfo{pages}{1893} (\bibinfo{year}{1975}); [\bibinfo{journal}{Sov. Phys. JETP} \textbf{\bibinfo{volume}{42}},
\bibinfo{pages}{961} (\bibinfo{year}{1976})].

\bibitem{Mitter}
\bibinfo{author}{\bibfnamefont{W.}~\bibnamefont{Becker}} \bibnamefont{and}
\bibinfo{author}{\bibfnamefont{H.}~\bibnamefont{Mitter}}.
\bibinfo{journal}{ J.\  Phys.\  A\ } \textbf{\bibinfo{volume}{8}},
\bibinfo{pages}{1638} (\bibinfo{year}{1975}).

\bibitem{Breit:1934zz}
\bibinfo{author}{\bibfnamefont{G.}~\bibnamefont{Breit}}
\bibnamefont{and}
\bibinfo{author}{\bibfnamefont{J.}~\bibfnamefont{A.}~\bibnamefont{Wheeler}}.
\bibinfo{journal}{Phys. Rev.}  \textbf{\bibinfo{volume}{46}}, \bibinfo{pages}{1087} (\bibinfo{year}{1934}).

\bibitem{Reiss1962}
\bibinfo{author}{\bibfnamefont{H.~R.}~\bibnamefont{Reiss}}.
\bibinfo{journal}{Jour. Math. Phys.}  \textbf{\bibinfo{volume}{3}}, \bibinfo{pages}{59} (\bibinfo{year}{1962}).

\bibitem{narozhnyi}
\bibinfo{author}{\bibfnamefont{N.~V.~} \bibnamefont{Narozhnyi}},
\bibinfo{author}{\bibfnamefont{A.~I.~} \bibnamefont{Nikishov}}
\bibnamefont{and}
\bibinfo{author}{\bibfnamefont{V.~I.~} \bibnamefont{Ritus}}.
[\bibinfo{journal}{Sov. Phys. JETP} \textbf{\bibinfo{volume}{20}},
\bibinfo{pages}{622} (\bibinfo{year}{1965})].

\bibitem{Titov}
\bibinfo{author}{\bibfnamefont{A.~I.}~\bibnamefont{Titov}},
\bibinfo{author}{\bibfnamefont{H.}~\bibnamefont{Takabe}},
\bibinfo{author}{\bibfnamefont{B.}~\bibnamefont{K\"ampfer}}
\bibnamefont{and}
\bibinfo{author}{\bibfnamefont{A.}~\bibnamefont{Hosaka}}.
\bibinfo{journal}{Phys.\  Rev.\ Lett. } \textbf{\bibinfo{volume}{108}},
\bibinfo{pages}{240406} (\bibinfo{year}{2012}).

\bibitem{krajewska}
\bibinfo{author}{\bibfnamefont{K.}~\bibnamefont{Krajewska}}
\bibnamefont{and}
\bibinfo{author}{\bibfnamefont{J.~Z.}~\bibnamefont{Kami\'nski}}.
\bibinfo{journal}{Phys.\  Rev.\ A. } \textbf{\bibinfo{volume}{86}},
\bibinfo{pages}{052104} (\bibinfo{year}{2012}).


\bibitem{VillalbaChavez:2012bb}
\bibinfo{author}{\bibfnamefont{S.}~\bibnamefont{Villalba-Chavez}}
\bibnamefont{and}
\bibinfo{author}{\bibfnamefont{C.} \bibnamefont{M\"uller}}.
\bibinfo{journal}{Phys. Lett. B},
\textbf{\bibinfo{volume}{718}},
\bibinfo{pages}{992}, \bibinfo{year}{2013}; arXiv:1208.3595 [hep-ph].

\bibitem{Milstein:2006zz}
\bibinfo{author}{\bibfnamefont{A.~I.}~\bibnamefont{Milstein}},
\bibinfo{author}{\bibfnamefont{C.}~\bibnamefont{M\"uller}},
\bibinfo{author}{\bibfnamefont{K.~Z.}~\bibnamefont{Hatsagortsyan}},
\bibinfo{author}{\bibfnamefont{U.~D.}~\bibnamefont{Jentschura}}
\bibnamefont{and}
\bibinfo{author}{\bibfnamefont{C.~H.}~\bibnamefont{Keitel}}.
\bibinfo{journal}{ Phys.\ Rev.\ A\ } \textbf{\bibinfo{volume}{73}},
\bibinfo{pages}{062106} (\bibinfo{year}{2006}).

\bibitem{Dipiazza:2010zz}
\bibinfo{author}{\bibfnamefont{A.~Di.}~\bibnamefont{Piazza}},
\bibinfo{author}{\bibfnamefont{E.}~\bibnamefont{L\"{otstedt}}},
\bibinfo{author}{\bibfnamefont{A.~I.}~\bibnamefont{Milstein}}
\bibnamefont{and}
\bibinfo{author}{\bibfnamefont{C.~H.}~\bibnamefont{Keitel}}.
\bibinfo{journal}{ Phys.\ Rev.\ A\ } \textbf{\bibinfo{volume}{81}},
\bibinfo{pages}{062122} (\bibinfo{year}{2010}).

\bibitem{Ahlers:2006iz}
\bibinfo{author}{\bibfnamefont{M.}~\bibnamefont{Ahler}},
\bibinfo{author}{\bibfnamefont{H.}~\bibnamefont{Gies}},
\bibinfo{author}{\bibfnamefont{J.}~\bibnamefont{Jaeckel}}  \bibnamefont{and}
\bibinfo{author}{\bibfnamefont{A.}~\bibnamefont{Ringwald}}.
\bibinfo{journal}{  Phys.\ Rev.\ D } \pmb{\bibinfo{volume}{75}},
\bibinfo{pages}{035011} (\bibinfo{year}{2007}); [hep-ph/0612098].

\bibitem{zavattini}
\bibinfo{author}{\bibfnamefont{G.}~\bibnamefont{Zavattini}}  \bibnamefont{and}
\bibinfo{author}{\bibfnamefont{E.}~\bibnamefont{Calloni}}.
\bibinfo{journal}{Eur.\  Phys. \ J.\ C } \pmb{\bibinfo{volume}{62}},
\bibinfo{pages}{459} (\bibinfo{year}{2009}).

\bibitem{Dyson:1949ha}
\bibinfo{author}{\bibfnamefont{F.~J.}~\bibnamefont{Dyson}}.
\bibinfo{journal}{Phys. Rev.},
\textbf{\bibinfo{volume}{75}},
\bibinfo{pages}{1736}, \bibinfo{year}{1949}.

\bibitem{Schwinger:1951ex1}
\bibinfo{author}{\bibfnamefont{J.~S.}~\bibnamefont{Schwinger}}.
\bibinfo{journal}{Proc. Nat. Acad. Sci.},
\textbf{\bibinfo{volume}{37}},
\bibinfo{pages}{452-455}, \bibinfo{year}{1951}.

\bibitem{Schwinger:1951ex2}
\bibinfo{author}{\bibfnamefont{J.~S.}~\bibnamefont{Schwinger}}.
\bibinfo{journal}{Proc. Nat. Acad. Sci.},
\textbf{\bibinfo{volume}{37}},
\bibinfo{pages}{455-459}, \bibinfo{year}{1951}.

\bibitem{fradkin}
\bibinfo{author}{\bibfnamefont{E.~S.}~\bibnamefont{Fradkin} in}
\bibinfo{journal}{Proceeding (Trudy) of the P. N. Lebedev Physics Institute, Vol.} \textbf{\bibinfo{volume}{29}},  (Consultants Bureau, New york, \bibinfo{year}{1967}).

\bibitem{Alkofer:2000wg}
\bibinfo{author}{\bibfnamefont{R.}~\bibnamefont{Alkofer}}
\bibnamefont{and}
\bibinfo{author}{\bibfnamefont{L.~von} \bibnamefont{Smekal}}.
\bibinfo{journal}{Phys. Rept.},
\textbf{\bibinfo{volume}{353}},
\bibinfo{pages}{281}, \bibinfo{year}{2001}.  [arXiv:hep-ph/0007355].

\bibitem{baierbook}
\bibinfo{author}{\bibfnamefont{V.~N.~} \bibnamefont{Ba\u{\i}er}},
\bibinfo{author}{\bibfnamefont{V.~M.~} \bibnamefont{Katkov}}
\bibnamefont{and}
\bibinfo{author}{\bibfnamefont{V.~M.~} \bibnamefont{Strakhovenko}}.
\emph{\bibinfo{book}{``Electromagnetic processes at high energies in oriented single crystals.''}}
\bibinfo{editor}{World Scientific}, Singapore, (\bibinfo{year}{1998}).

\bibitem{sophocles}
\bibinfo{author}{\bibfnamefont{S.~J.~} \bibnamefont{Orfanidis}}
\bibnamefont{in}
\emph{\bibinfo{book}{``Electromagnetic Waves and Antennas.''}}
\bibinfo{editor}{Chap 4}, online-book\\
http://www.ece.rutgers.edu/~orfanidi/ewa/

\bibitem{spin}
\bibinfo{author}{\bibfnamefont{S.}~\bibnamefont{Ahrens}},
\bibinfo{author}{\bibfnamefont{T.~O.}~\bibnamefont{M\"{u}ller}},
\bibinfo{author}{\bibfnamefont{S.}~\bibnamefont{Villalba-Chavez}},
\bibinfo{author}{\bibfnamefont{H.}~\bibnamefont{Bauke}} \bibnamefont{and}
\bibinfo{author}{\bibfnamefont{C.} \bibnamefont{M\"{u}ller}}.
\bibinfo{journal}{J.\ Phys.:\  Conf. \ Ser.\ } \textbf{\bibinfo{volume}{414}},
\bibinfo{pages}{012012}, \bibinfo{year}{2013}.

\bibitem{Richard}
\bibinfo{author}{\bibfnamefont{J.~L.}~\bibnamefont{Richard}}.
\bibinfo{journal}{Nuovo Cimento}  \textbf{\bibinfo{volume}{8A}}, \bibinfo{pages}{485} (\bibinfo{year}{1972}).

\bibitem{Heinzl}
\bibinfo{author}{\bibfnamefont{T.~} \bibnamefont{Heinzl}},
\bibinfo{author}{\bibfnamefont{B.~} \bibnamefont{Leifeld}},
\bibinfo{author}{\bibfnamefont{K.~U.} \bibnamefont{Amthor}},
\bibinfo{author}{\bibfnamefont{H.~} \bibnamefont{Schwoerer}},
\bibinfo{author}{\bibfnamefont{R.~} \bibnamefont{Sauerbrey}}
\bibnamefont{and}
\bibinfo{author}{\bibfnamefont{A.~} \bibnamefont{Wipf}}.
\bibinfo{journal}{Opt. Comm.} \pmb{\bibinfo{volume}{267}},
\bibinfo{pages}{318} (\bibinfo{year}{2006}).

\bibitem{axion}
\bibinfo{author}{\bibfnamefont{S.}~\bibnamefont{Villalba-Chavez}}
in
\emph{\bibinfo{title}{``Laser-driven search of axion-like particles  including  vacuum polarization  effects.''}}; arXiv:1308.4033 [hep-ph].

\bibitem{DiPiazza:2006pr}
  A.~Di Piazza, K.~Z.~Hatsagortsyan and C.~H.~Keitel,
  Phys.\ Rev.\ Lett.\  {\bf 97} (2006) 083603
  [hep-ph/0602039].

\bibitem{Gradshteyn}
\bibinfo{author}{\bibfnamefont{I.}~\bibfnamefont{S.}~\bibnamefont{Gradshteyn}}
\bibnamefont{and}
\bibinfo{author}{\bibfnamefont{I.}~\bibfnamefont{M.}~\bibnamefont{Ryzhik}}.
\emph{\bibinfo{book}{``Table of Integrals, Series and Products.''}}
\textrm{Seventh Edition}, \bibinfo{editor}{Elsevier}, San Diego, (\bibinfo{year}{2007}).

\bibitem{Olver}
\bibinfo{author}{\bibfnamefont{F.~W.~J.}~\bibnamefont{Olver}}.
\emph{\bibinfo{book}{``Asymptotics and special functions.''}}
\textrm{Tenth Printing}, \bibinfo{editor}{Academic Press}, London, (\bibinfo{year}{1974}).

\bibitem{abramowitz}
\bibinfo{author}{\bibfnamefont{M.}~\bibnamefont{Abramowitz}}
\bibnamefont{and}
\bibinfo{author}{\bibfnamefont{I.}~\bibfnamefont{A.}~\bibnamefont{Stegun}}.
\emph{\bibinfo{book}{``Handbook of Mathematical Functions.''}}
\textrm{Tenth Printing}, \bibinfo{editor}{National Bureau of Standards}, USA, (\bibinfo{year}{1973}).

\bibitem{Katkov}
\bibinfo{author}{\bibfnamefont{V.~N.~} \bibnamefont{Ba\u{\i}er}},
\bibnamefont{and}
\bibinfo{author}{\bibfnamefont{V.~M.~} \bibnamefont{Katkov}}.
\bibinfo{journal}{Nucl. Instrum. Meth. B} \pmb{\bibinfo{volume}{243}},
\bibinfo{pages}{282} (\bibinfo{year}{2006}).

\bibitem{ritus}
\bibinfo{author}{\bibfnamefont{V.~I.}~\bibnamefont{Ritus}}.
\bibinfo{journal}{Ann.  Phys.} \textbf{\bibinfo{volume}{69}},
\bibinfo{pages}{55} (\bibinfo{year}{1972}).

\bibitem{PHELIX} see: https://www.gsi.de/en/start/research/forschungsgebiete\_und\_experimente/appa\_pni\_gesundheit/plasma\_physicsphelix/phelix.htm

\bibitem{Muroo_2003}
\bibinfo{author}{\bibfnamefont{K.~} \bibnamefont{Muroo}  {\it et al.}}
\bibinfo{journal}{J.\  Opt.\ Soc.\  Am.\  B\ } \pmb{\bibinfo{volume}{20}},
\bibinfo{pages}{2249} (\bibinfo{year}{2003}).

\bibitem{POLARIS}
\bibinfo{author}{\bibfnamefont{M.~} \bibnamefont{Hornung}  {\it et al.}}
\bibinfo{journal}{Appl. Phys. B} \pmb{\bibinfo{volume}{101}},
\bibinfo{pages}{93} (\bibinfo{year}{2010}).

\bibitem{Dobrich:2010hi}
\bibinfo{author}{\bibfnamefont{B.}~\bibnamefont{D\"obrich}} \bibnamefont{and}
\bibinfo{author}{\bibfnamefont{H.}~\bibnamefont{Gies}}.
 \bibinfo{journal}{JHEP} \textbf{\bibinfo{volume}{1010}},
\bibinfo{pages}{022} (\bibinfo{year}{2010}).

\bibitem{Dobrich:2010ie}
\bibinfo{author}{\bibfnamefont{B.}~\bibnamefont{D\"obrich}} \bibnamefont{and} \bibinfo{author}{\bibfnamefont{H.}~\bibnamefont{Gies}}.
\emph{``High-Intensity Probes of Axion-Like Particles,''}
Contributed to 6th Patras Workshop on Axions, WIMPs and WISPs, Zurich, Switzerland, 5-9 Jul 2010.

\bibitem{Burke:1997ew}
\bibinfo{author}{\bibfnamefont{D.~L.}~\bibnamefont{Burke}}
\bibnamefont{et~al.}
\bibinfo{journal}{Phys. Rev. Lett.}  \textbf{\bibinfo{volume}{79}}, \bibinfo{pages}{1626} (\bibinfo{year}{1997}).





\end{thebibliography}
 \end{document}